\pdfoutput=1
\documentclass[12pt,oneside,english]{amsart}

\usepackage[T1]{fontenc}
\usepackage[latin9]{inputenc}
\usepackage{geometry}
\usepackage{babel}
\usepackage{amsbsy}
\usepackage{amstext}
\usepackage{amsthm}
\usepackage{amssymb}
\usepackage{graphicx}
\usepackage[unicode=true,pdfusetitle,
 bookmarks=true,bookmarksnumbered=false,bookmarksopen=false,
 breaklinks=false,pdfborder={0 0 1},backref=false,colorlinks=false]
 {hyperref}

\makeatletter

\newcommand{\lyxdot}{.}

\numberwithin{equation}{section}
\numberwithin{figure}{section}

\makeatother

\begin{document}

\title{Bulk Spectrum and $K$-theory for Infinite-Area Topological Quasicrystals}

\author{Terry A. Loring}

\address{Department of Mathematics and Statistics, University of New Mexico,
Albuquerque, New Mexico 87131, USA}

\keywords{Chern insulator, bulk spectrum, quasilattice, numerical linear algebra,
Bott index.}
\begin{abstract}
The bulk spectrum of a possible Chern insulator on a quasicrystalline lattice is examined. The effect of being a 2D insulator seems to override any fractal properties in the spectrum. We compute that the spectrum is either two continuous bands, or that any gaps other than the main gap are small. After making estimates on the spectrum, we deduce a finite system size, above which the $K$-theory must coincide with the $K$-theory of the infinite system. Knowledge of the spectrum and $K$-theory of the infinite-area system will control the spectrum and $K$-theory of sufficiently large finite systems.

The relation between finite volume $K$-theory and infinite volume Chern numbers is only proven to begin, for the model under investigation here, for systems on Hilbert space of dimension around 17 million.  The real-space method based on the Clifford spectrum allows for computing Chern numbers for systems on Hilbert space of dimension around $2.7$ million. New techniques in numerical $K$-theory are used to equate the $K$-theory of systems of different sizes.
\end{abstract}

\maketitle
\markright {Bulk Spectrum and $K$-theory for Infinite-Area Topological Quasicrystals}\markleft {}

\section{Introduction}

Recent research \cite{bandres2016topological,fulga2016aperiodic,HuangLiu_spin_Bott,tran2015topological}
shows that topological insulators and superconductors can occur as
a quasicrystal. There is $K$-theory \cite{bourne2018non,HuangLiu_QSH_quasilattice,HuangLiu_spin_Bott}
associated to gaps in the bulk spectrum, and edge states \cite{fulga2016aperiodic,HuangLiu_QSH_quasilattice,tran2015topological}
that arise within these gaps at the boundaries of finite systems.
In the crystalline case, we have little doubt as to how to define
and calculate the bulk spectrum, so most of the discussion has been
on how to detect and calculate $K$-theory.

Let us step away from the $K$-theory for a while, and ponder: what
is the bulk spectrum for the Hamiltonian in a tight-binding model
on a quasicrystalline lattice, and how do we compute it? 

A noncontroversial means to define bulk spectrum is simply to use
the spectrum of the Hamiltonian for the system on full, infinite-area
quasicrystalline lattice. The issue is how is a physicist to compute
this? We have lost translation invariance, so cannot use momentum
space to reduce the calculation to finding the spectrum of a parameterized
collection of small matrices. Perhaps also gone is approximating the
bulk spectrum by the spectrum of a finite sample with periodic boundary
conditions. 

One can impose periodic boundary conditions on a periodic approximation
to the quasilattice, at the cost of introducing a few defects \cite{bandres2016topological,tsunetsugu1991electronic}.
It is not clear exactly the relation between the spectrum of these
finite systems and that of the infinite system. Perhaps more theoretical
work using more explicitly $C^{*}$-algebras, such as in \cite{beckus2016continuity,beckus2017spectral,prodan2016bulk},
can settle this issue.

Another method to investigate the bulk spectrum of a quasiperiodic
Hamiltonian is to look at the effect of self-similarity. This method
was applied \cite{sire1990renormalization} to a Hamiltonian different
from the one studied here, but also on the Ammann-Beenker tiling.
This system had no topological index to confound the picture and the
conclusion on the bulk spectrum was very different from what is presented
here.

Wanting to know the spectrum of the infinite system is more than a
fascinating mathematical question. It provides a manner to predict
the behavior of finite systems of arbitrarily large size. We hope
that the edge phenomena seen in specific finite systems of one or
two sizes will persist in larger systems. However, we shall see that
there are edge phenomena associated to apparent small gaps in ``the
bulk spectrum'' of modest-sized systems and we can not easily tell
if these gaps will disappear when we look at larger systems.

We will focus on a single infinite-area Hamiltonian because the numerical
computations of the bulk spectrum are somewhat demanding. Specifically
we investigate the ``$p_{x}+ip_{y}$'' tight-binding model on the
Ammann-Beenker tiling, introduced in previous work with Fulga and
Pikulin  \cite{fulga2016aperiodic}. After numerically estimating
the bulk spectrum, we are able
to calculate a size above which all finite models have the same $K$-theory
as the infinite model. We find an indirect way to calculate the $K$-theory
associated to one such infinite model, thus providing solid evidence
that the infinite system has Chern number $-1$. 

The methods introduced do not depend on the structure of the quasilattice.
They can be implemented whenever the Hamiltonian is local and bounded,
although the accuracy and efficiency are expected to vary according
to the structure of the quasilattice.

\section{A tight-binding Hamiltonian on a Quasicrystal}

\begin{figure}
\includegraphics[width=4in]{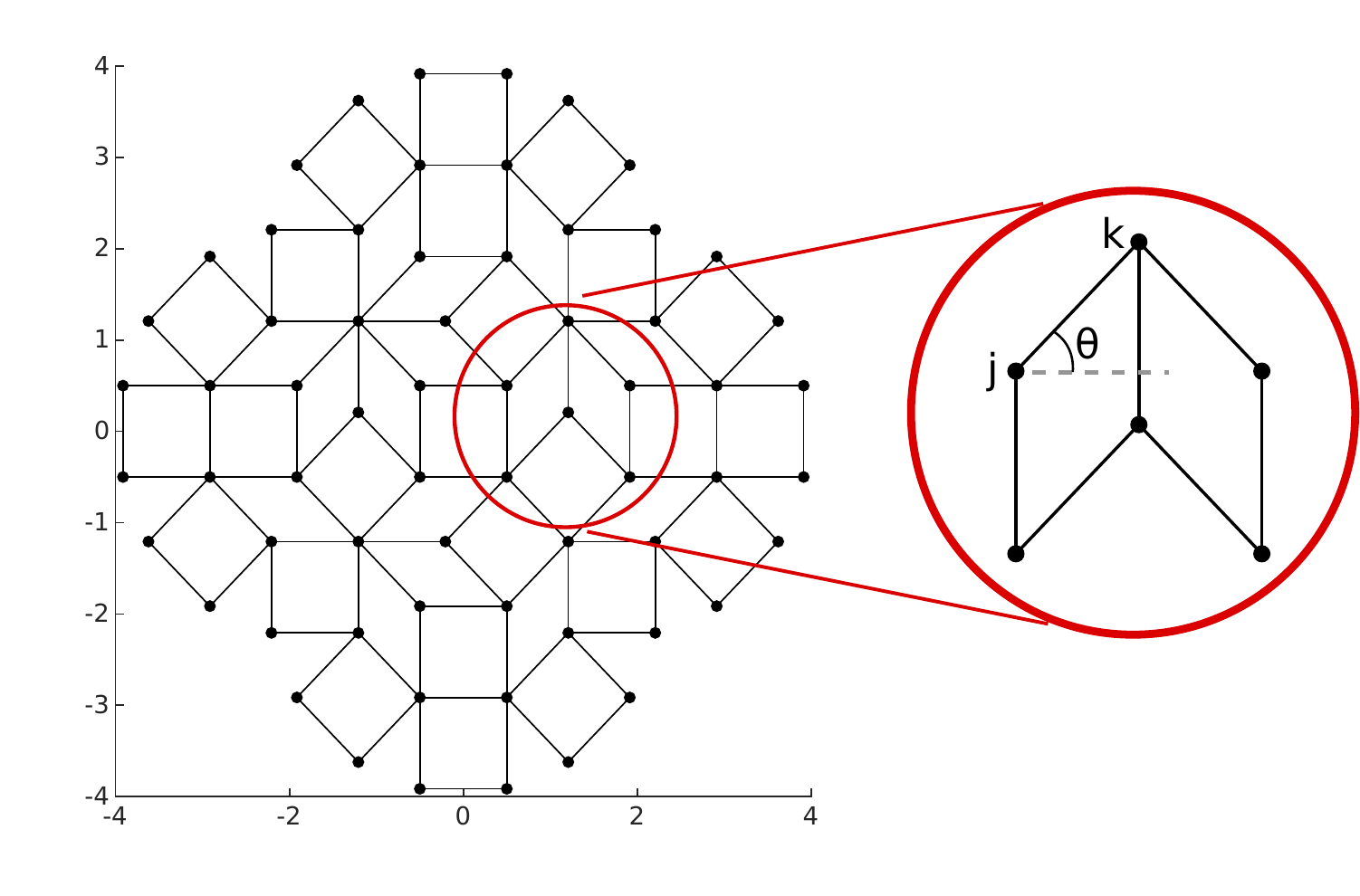}

\caption{Defining the infinite-area Hamiltonian $H_{\mathrm{QC}}$, determined
by the quasilattice. Shown is the definition of the angles $\theta=\theta_{jk}$
associated to the $j$-$k$ term. We define the lattice unit to be
the common length of all the edges. \label{fig:Defining-infinite-Hamiltonian}}

\end{figure}

We will investigate the ``$p_{x}+ip_{y}$'' tight-binding model on
the Ammann-Beenker tiling \cite{beenker1982algebraic}. A small system
using this Hamiltonian was examined previously \cite{fulga2016aperiodic}.
The Hilbert space is $\ell^{2}(\mathcal{V})\otimes\mathbb{C}^{2}$,
where $\mathcal{G}=(\mathcal{E},\mathcal{V})$ is the graph whose
edges and vertices are taken from the tiles of an Ammann-Beenker tiling
of the plane. Define $H_{\mathrm{QC}}$, a bounded Hermitian operator
on $\ell^{2}(\mathcal{V})\otimes\mathbb{C}^{2}$, as an infinite matrix
with 
\[
H_{j}=-\mu\sigma_{z},
\]
 a term associated with each site (vertex), and 
\[
H_{jk}=-t\sigma_{z}-\frac{i}{2}\Delta\sigma_{x}\cos(\theta_{jk})-\frac{i}{2}\Delta\sigma_{y}\sin(\theta_{jk}),
\]
 a hopping term associated to each edge. The angle is taken from a
reference angle, as illustrated in Figure~\ref{fig:Defining-infinite-Hamiltonian}.

\section{Working around Edge States}

\begin{figure}
\includegraphics[bb=120bp 200bp 490bp 590bp,clip,width=4in]{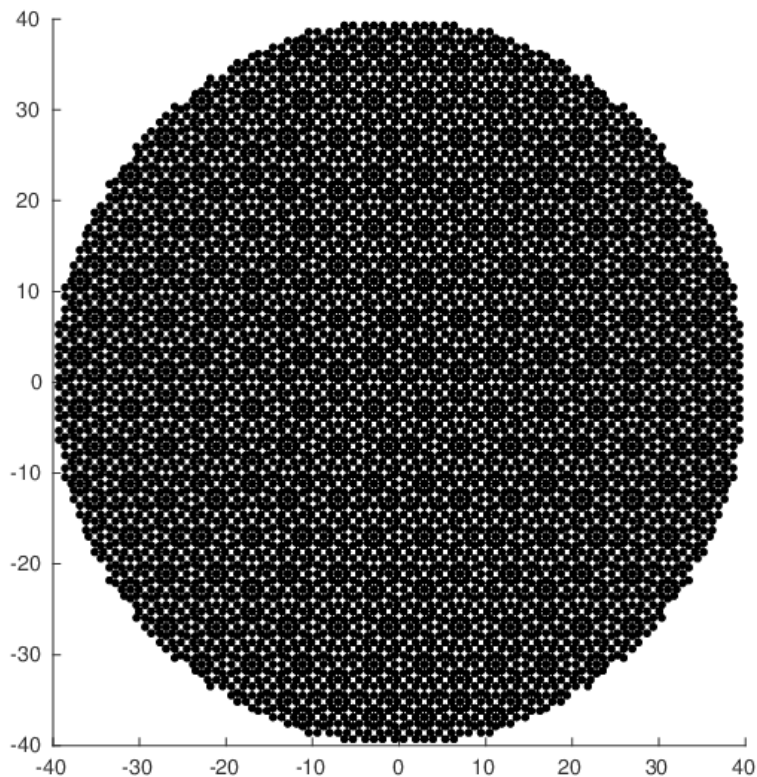}

\caption{Round section of the quasilattice, of radius 40 lattice units. \label{fig:Round-Lattice}}

\end{figure}

A direct approach to estimating the bulk spectrum is to put Dirichlet
boundary conditions on a finite sample, compute the spectrum of the
finite Hamiltonian, and sort those eigenvalues according to whether
the associated eigenvector seems to be localized at the edge or not.
This technique has certainly been used before \cite{bandres2016topological}
for this purpose. We prefer to use round samples, as theory \cite{loringSB_even_dim_localizer}
suggests that the smallest distance to the center of a sample is what
dictates the strength of small system size effects. Also, this was
the choice made in a earlier study \cite{repetowicz1998exact} of
an ordinary insulator on a quasilattice. A typical round portion of
the quasilattice is shown in Figure~\ref{fig:Round-Lattice}. 

For a given $\rho$, we will denote the Hamiltonian for the finite
system on the round portion of the lattice centered at the origin
of radius $\rho$ by $H_{\rho}$. We look also at Hamiltonians and
round finite systems at other locations, but find these of little
additional benefit and do not introduce notation for them.

We begin with an easy calculation, finding the norm of $H_{QC}$, which
is just the largest eigenvalue. We remark that $H_{QC}$ has a form
of particle-hole symmetry that is not of much interest here. It
does tells us that all our eigenvalue calculations will be symmetric under
the transformation $\lambda\mapsto-\lambda$. We will thus be able
to focus on the eigenvalues that are zero or positive. It also
allows us to find a unitary change of basis in some situations to
transform a complex matrix into a real matrix. This leads to being
able to work with slightly larger systems with the same computing
hardware. 

\begin{figure}
\includegraphics[bb=160bp 320bp 430bp 465bp,clip,width=4in]{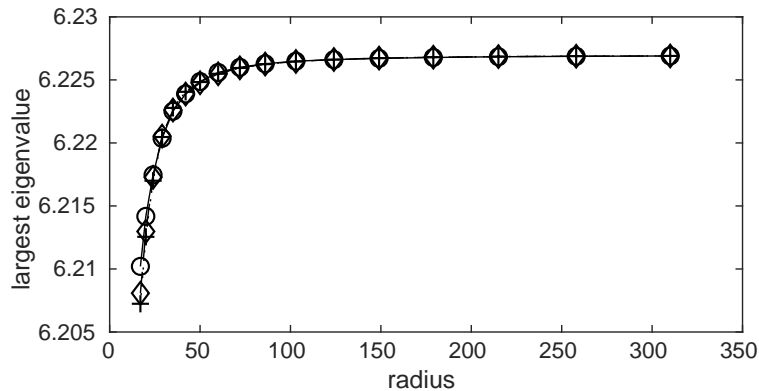}

\caption{Norms of the compressed Hamiltonians $H_{\rho}$ as an estimator of
the norm of $H_{QC}$ (solid line, round markers). Also computed for
round portions of the same radii but centered at $(300,0)$ (dotted
line, with plus markers) and also centered at $(-100,-200)$ (dashed
line, with diamond markers). \label{fig:Norms-various-radii}}

\end{figure}

The estimates on $\|H_{QC}\|$ look very similar if we locate the
centers of the round portions at different locations. A manifestation
of quasiperiodicity seems to be that for most calculations it matters
little from where in the infinite quasilattice our finite lattices
are taken. The conclusion we can make from the data in Figure~\ref{fig:Norms-various-radii}
is $\|H_{QC}\|\approx6.227$.

\begin{figure}
\includegraphics[bb=0bp 25bp 244bp 140bp,clip,width=6in]{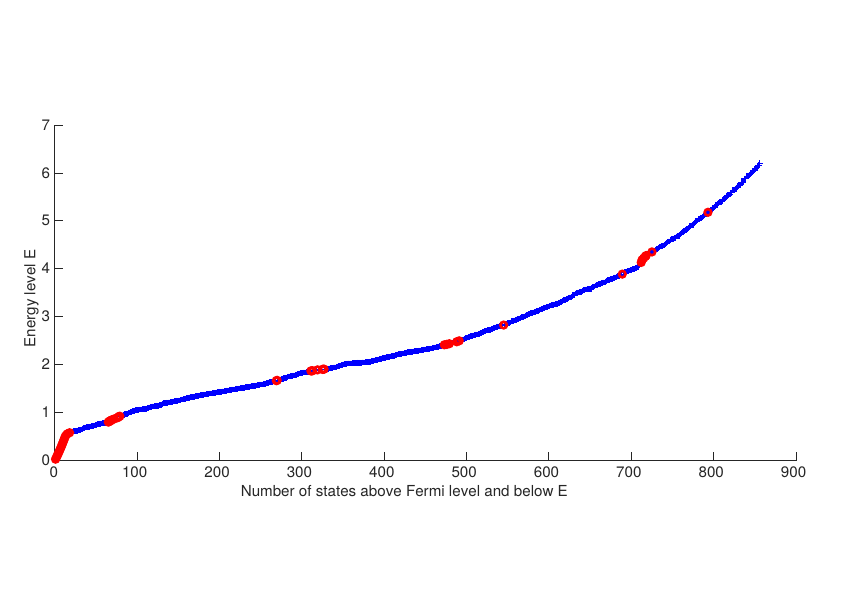}\\
\includegraphics[bb=0bp 5bp 244bp 120bp,clip,width=6in]{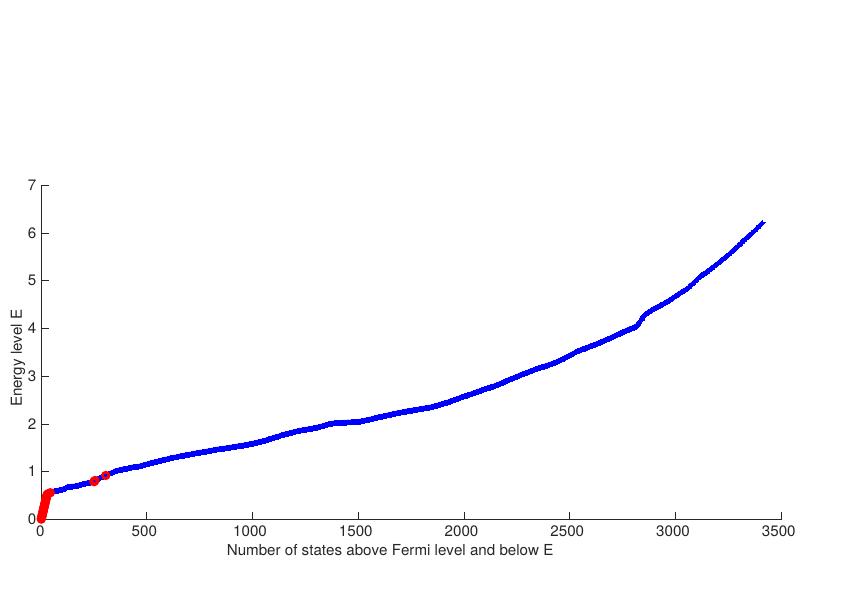}

\caption{Integrated density of states, positive energy only. Each energy level
is marked by a large, light, red circle if it corresponds to what
seems to be a boundary mode, a small, dark, blue cross if corresponding
to a bulk mode. Boundary mode is defined, rather arbitrarily, to correspond
to the expectation of observing this mode within 3 lattice units of the boundary circle being at least $1/2$. Top panel is for radius 15. Bottom panel
is for radius 25. \label{fig:IDOS-small-systems}}

\end{figure}

The expected edge states induced by the nonzero topological index
will make finding the smallest eigenvalue more difficult than finding
the largest eigenvalue. One strategy is to sort the eigenvalues into
``bulk'' and ``edge'' states by examining the associated states (eigenvectors).
We arbitrarily declare a state to be an edge state if, according to
its probability distribution, it is more than 50\% likely to be observed
to be outside of a circle of radius $\rho-3$, where $\rho$ is the
radius of the finite system. We compute the integrated density of
states, for positive energy only, and find many possible gaps beyond
the expected gap at zero when looking at a small system of radius
$\rho=15$. For $\rho=25$ there are still some small runs of edge
states. See Figure~\ref{fig:IDOS-small-systems}.

\begin{figure}
\includegraphics[bb=85bp 300bp 520bp 495bp,clip,width=6in]{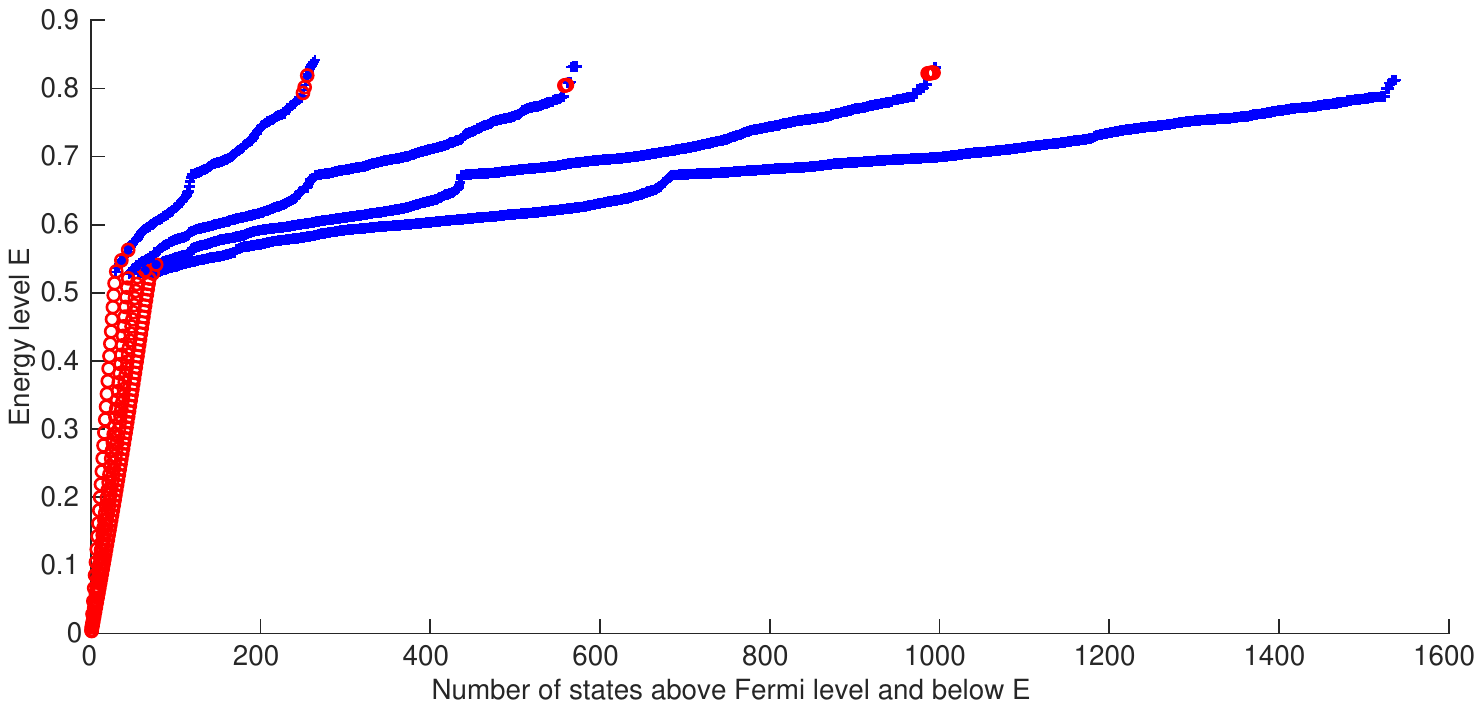}

\caption{A portion of the IDOS. Radii 30, 45, 60 and 75 shown together. \label{fig:IDOS-larger}}

\end{figure}

Looking at somewhat larger systems, up to $\rho=75$, we see in Figure~\ref{fig:IDOS-larger}
that the possible gap in the bulk spectrum near $E=0.8$ fades away.
This is not a rigorous argument, and we can't yet rule out small gaps
in the bulk coming back at larger system size.

\begin{figure}
\includegraphics[bb=180bp 270bp 430bp 525bp,clip,width=2.6in]{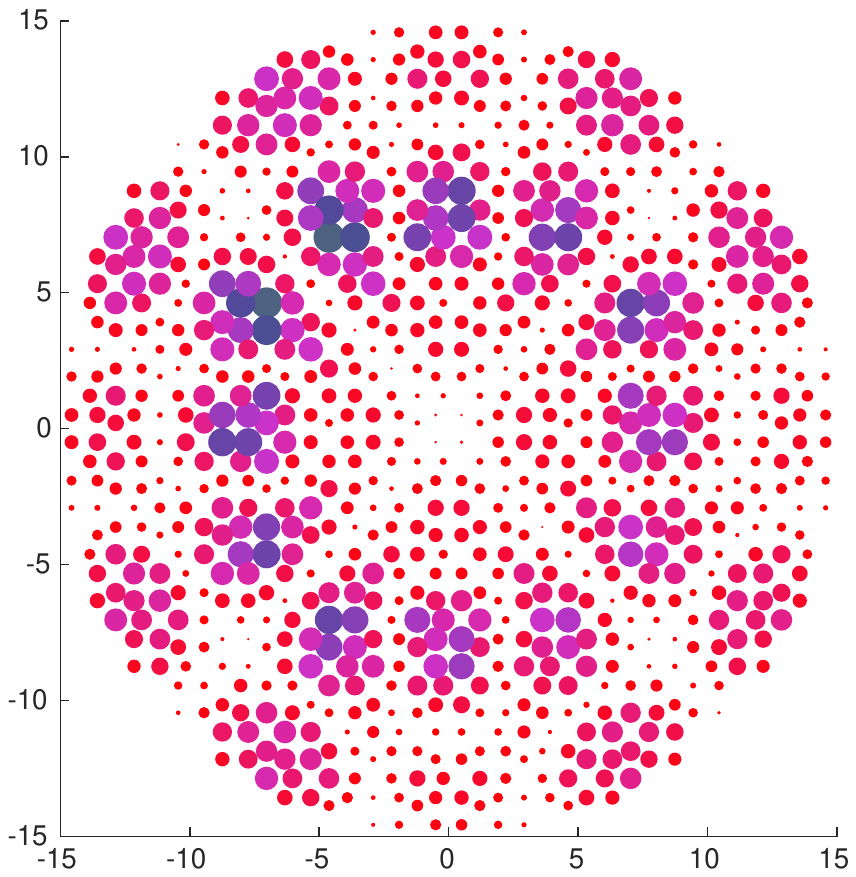}\includegraphics[bb=180bp 270bp 430bp 525bp,clip,width=2.6in]{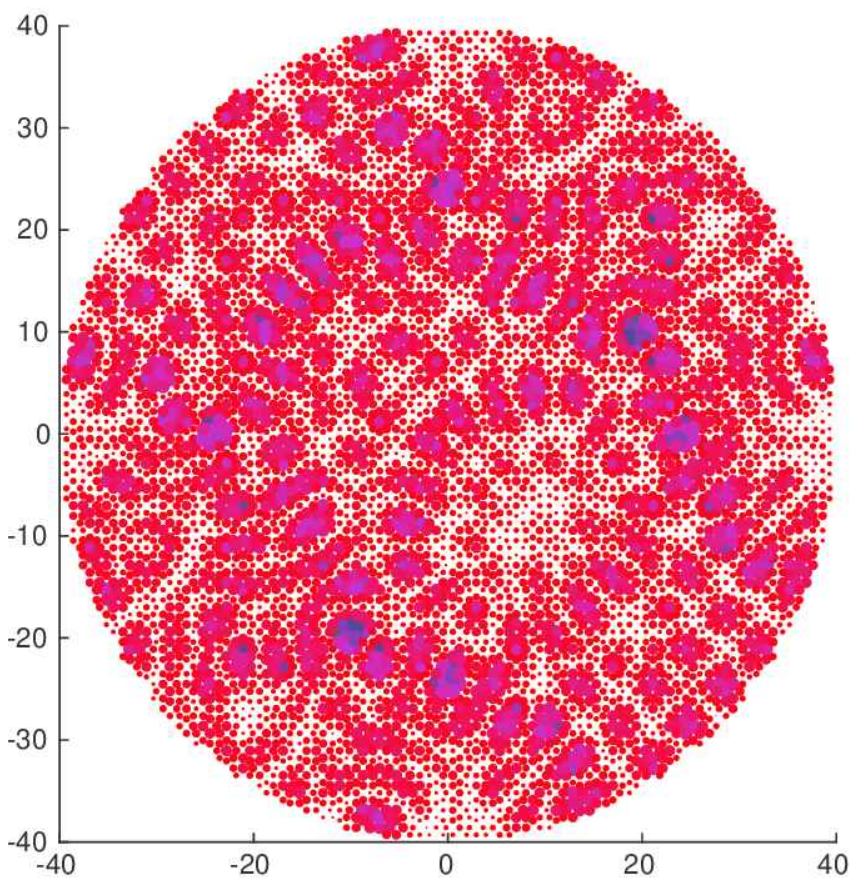}

\caption{On the left the radius is 15, with eigenvalue $5.5066$. On the right
the radius is 40 with eigenvalue $5.498$. Blue and dark dots are
holes while red and lighter dots are particles. The area of each dot
corresponds to the probability that this state is measured to be at
that location. \label{fig:bulk-states}}

\end{figure}

\begin{figure}
\includegraphics[bb=180bp 270bp 430bp 525bp,clip,width=2.6in]{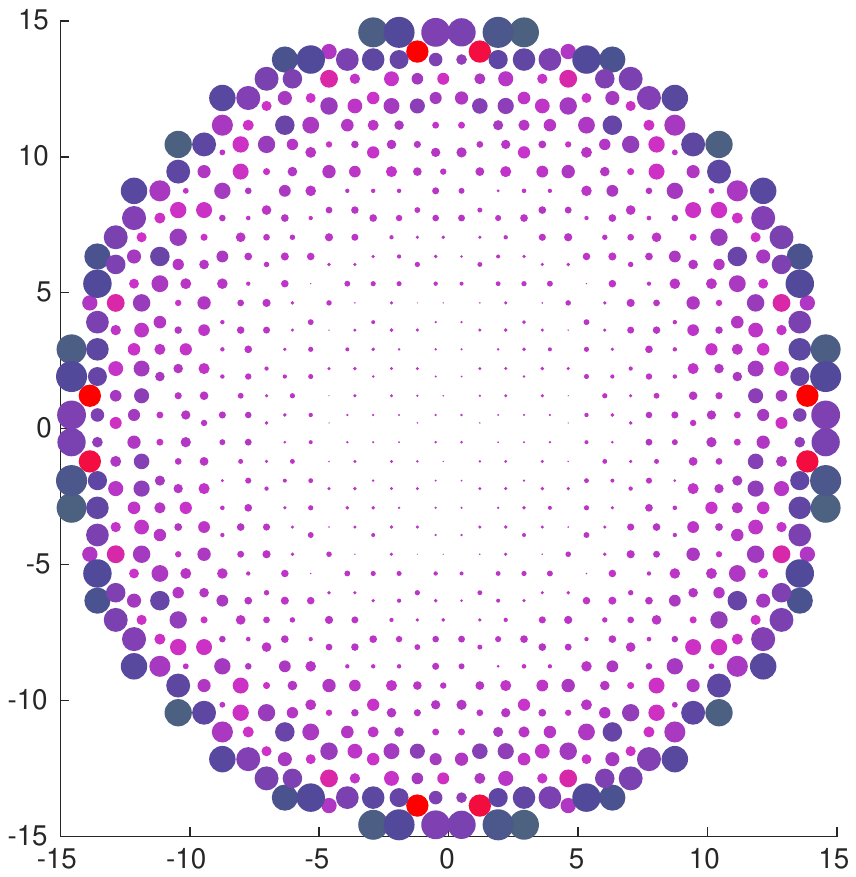}\includegraphics[bb=180bp 270bp 430bp 525bp,clip,width=2.6in]{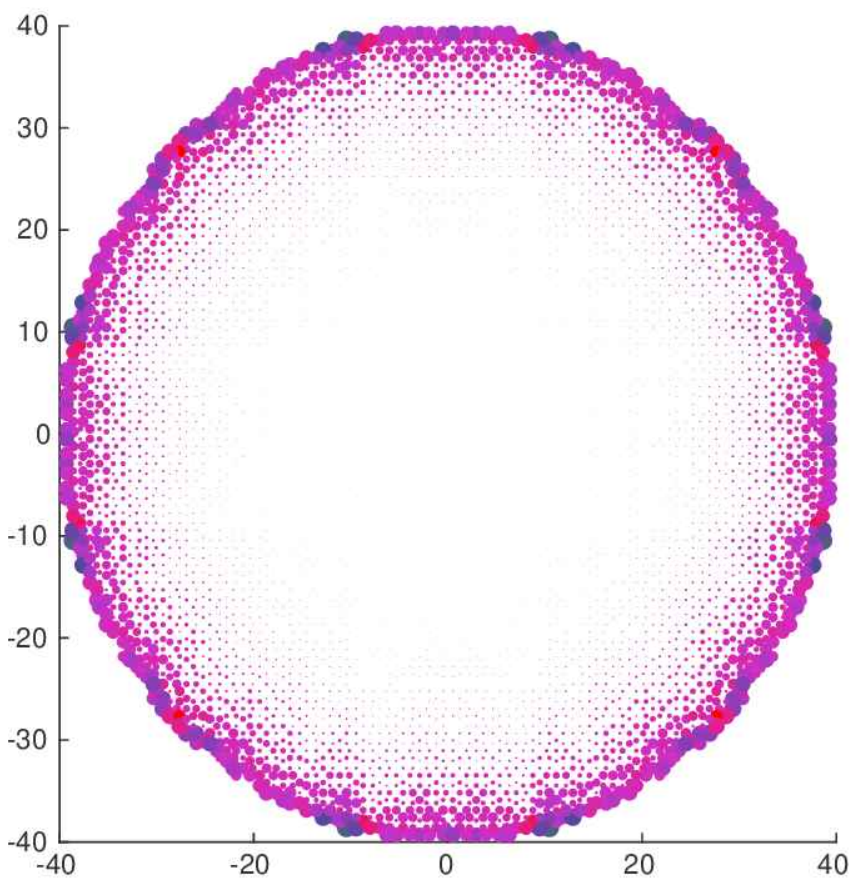}

\caption{On the left the radius is 15, with eigenvalue $0.018$. On the right
the radius is 40 with eigenvalue $0.007$. \label{fig:edge-states}}
\end{figure}

States that are decidedly ``bulk states'' occur at many energies and
system sizes. Two are illustrated in Figure~\ref{fig:bulk-states}.
States that are decidedly ``edge states'' are common near zero energy,
and two of these are illustrated in Figure~\ref{fig:edge-states}.

\begin{figure}
\includegraphics[bb=180bp 270bp 430bp 525bp,clip,width=2.6in]{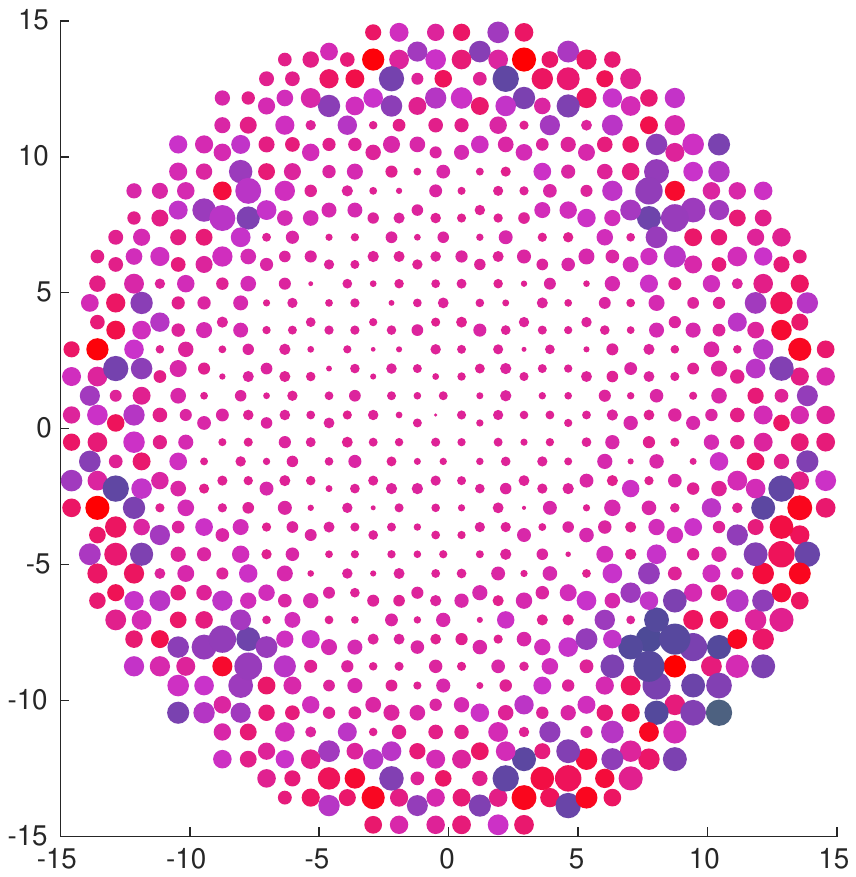}\includegraphics[bb=180bp 270bp 430bp 525bp,clip,width=2.6in]{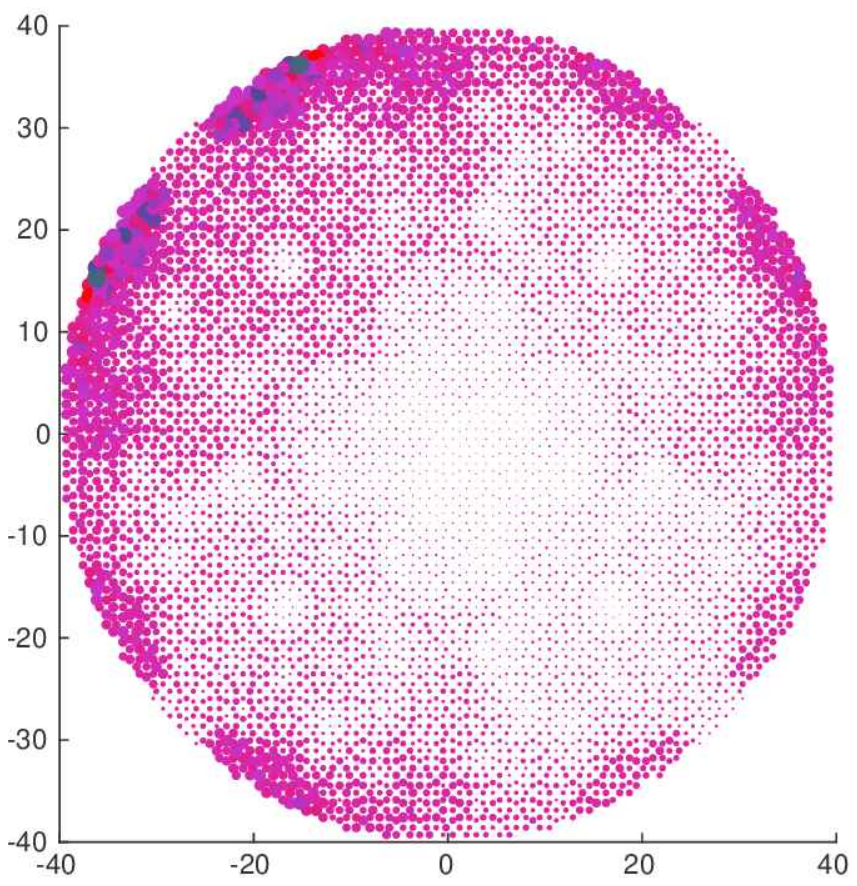}

\caption{On the left the radius is 15, with eigenvalue $0.803$. On the right
the radius is 40 with eigenvalue $0.828$. \label{fig:ambiguous_mode}}
\end{figure}

What makes the situation complicated is the existence of states like
those in Figure~\ref{fig:ambiguous_mode}. At energy around $0.8$
we see some modes that are not really edge states and not really bulk
states.

\begin{figure}
\includegraphics[bb=180bp 270bp 430bp 525bp,clip,width=2.6in]{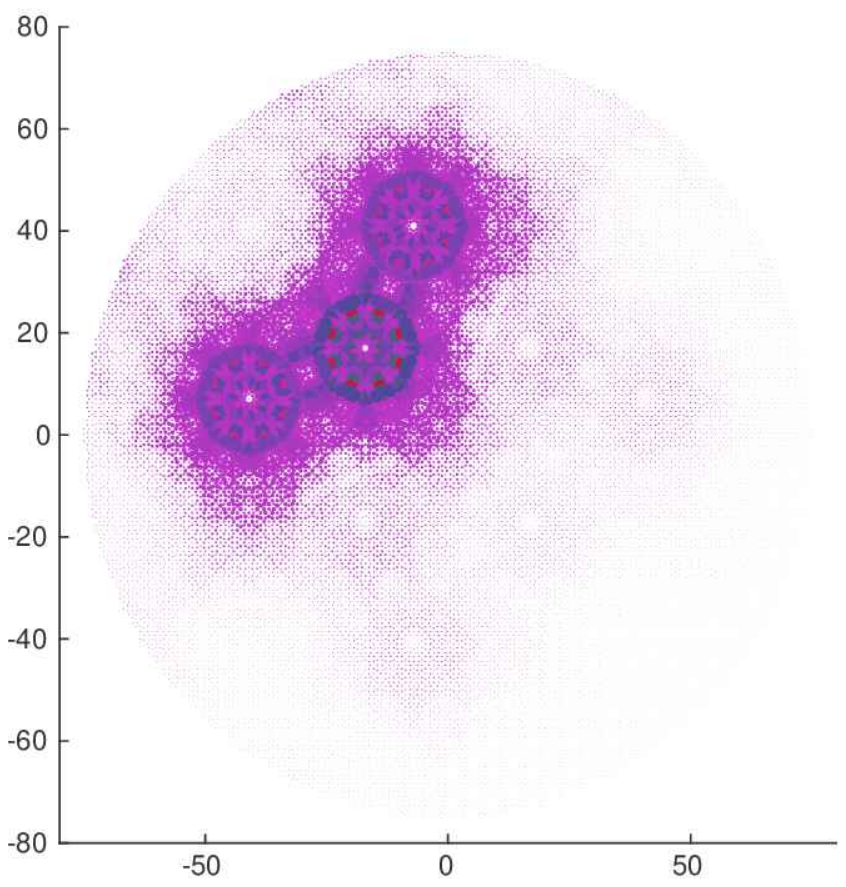}\includegraphics[bb=180bp 270bp 430bp 525bp,clip,width=2.6in]{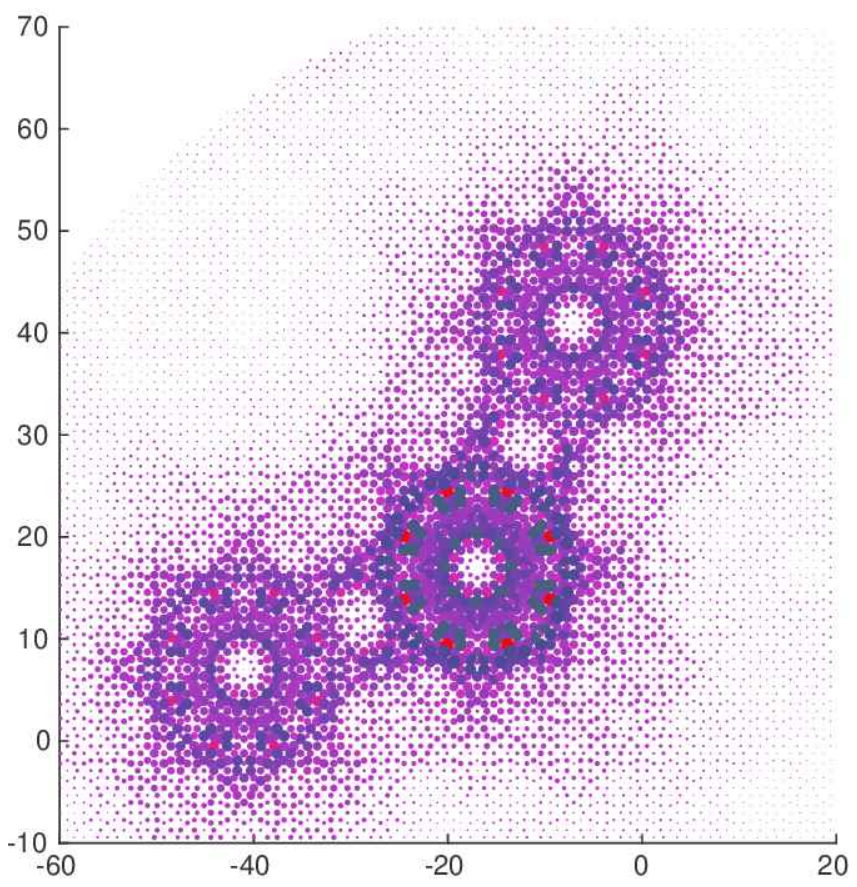}

\caption{The eigenvector corresponding to $\lambda=0.831$ with radius 75.
On the left the entire distribution. On the right, zoomed in to the
densest portion. \label{fig:bigger_system_less_ambiguous}}
\end{figure}

\begin{figure}
\includegraphics[bb=115bp 210bp 495bp 590bp,clip,width=4in]{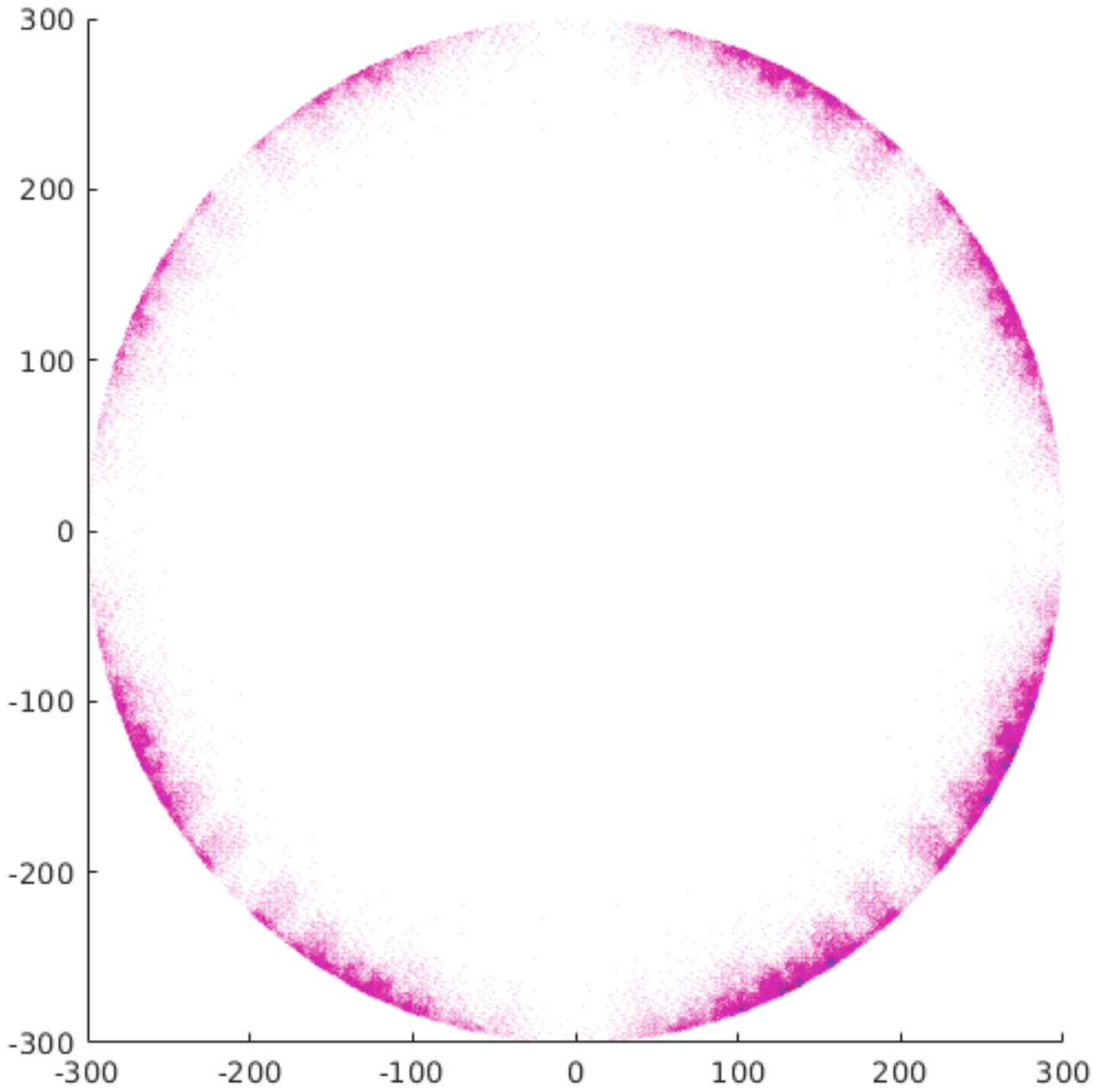}

\caption{The eigenvector corresponding to $\lambda=0.828$ with radius $\rho=300$
is again ambiguous, a state somewhat on the edge. \label{fig:BigAndAmbiguous}}
\end{figure}

One way to avoid such an ambiguous state is to work with a larger
system, as in Figure~\ref{fig:bigger_system_less_ambiguous}. Unfortunately,
we again find ambiguity with a yet larger system, as shown in Figure~\ref{fig:BigAndAmbiguous}.

\section{Approximate Eigenvectors}

To be more rigorous, we turn away from exact eigenvectors of a small
Hamiltonian $H_{\rho}$ and investigate instead approximate eigenvectors
of the full $H_{QC}$. Fortunately, the two concepts are related.
As $H_{QC}$ is Hermitian, its spectrum equals its approximate point
spectrum \cite{demuth2005selfadjointness}. Moreover, every vector
is close to a vector with support within a large circle at the origin.
We say a vector $\psi$ on the Hilbert space for $H_{\rho}$ is \emph{zero
at the boundary} if it has value zero on all canonical basis element
within one lattice unit of the bounding circle of radius $\rho$.
If we pad $\psi$ with zeros, then the fact that $H_{\rho}$ has no
hopping terms acting over a length more than $1$ means that 
\[
H_{QC}\left[\begin{array}{c}
\psi\\
0
\end{array}\right]=\left[\begin{array}{c}
H_{\rho}\psi\\
0
\end{array}\right].
\]
 All these facts constitute a proof that $\lambda$ is in the spectrum
of $H_{QC}$ if, and only if, there is a sequence of vector states $\psi_{n}$
that are zero at the boundary in the truncated Hilbert space of radius
$\rho_{n}$ such that 
\[
\lim_{n\rightarrow\infty}\left\Vert H_{\rho_{n}}\psi_{n}-\lambda\psi_{n}\right\Vert =0.
\]

A more useful estimate can be obtain using Weyl's estimate (see Bhatia's
book \cite{bhatia2013matrix} for example) relating the spectrum of
a perturbed Hermitian operator and the size of the perturbation. If
we can produce a single state $\psi$ in the Hilbert space of radius
$\rho_{n}$ that is zero at the boundary with 
\[
\left\Vert H_{\rho}\psi-\lambda\psi\right\Vert \leq\epsilon
\]
 then there is a point in the spectrum of $H_{QC}$ within distance
$\epsilon$ of $\lambda$.

Given an eigenstate $\psi$ for $H_{\rho}$ with eigenvalue $\lambda$,
we can multiply it by a tapering function, rescale the result to obtain
$\psi'$, and calculate the ``eigenerror'' $\left\Vert H_{\rho}\psi'-\lambda\psi'\right\Vert $
which is also a bound on the distance to the spectrum of the operator
we care about, $H_{QC}$. This gets around the artificial dichotomy
between bulk and edge states. Instead, we consider a state to be more
concentrated in the bulk when the resulting eigenerror is smaller.
This measure on bulkiness depends, of course, on our choice of taper
function. We will use a radial function, where the scalar $\tau(r)$
used depends on the radius $r$ from the system center via 
\begin{equation}
\tau(r)=\begin{cases}
1-\frac{2}{\rho_{1}^{2}}r{}^{2} & r<\frac{\rho_{1}}{2}\\
1-\frac{2}{\rho_{1}^{2}}(r-\rho_{1})^{2} & \frac{\rho_{1}}{2}\leq r\leq\rho_{1}\\
0 & \rho_{1}<r
\end{cases}\label{eq:taper_function}
\end{equation}
 where $\rho_{1}=\rho-1$.

\begin{figure}
\includegraphics[bb=50bp 325bp 580bp 465bp,clip,scale=0.7]{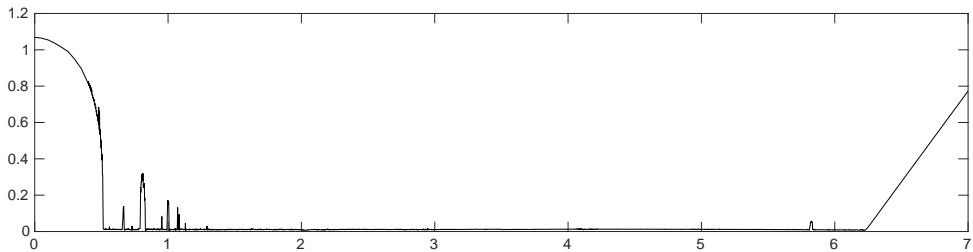}

\caption{Upper bound on the distance to the bulk spectrum via approximate eigenvalues
that were found by the compression method. \label{fig:Upper-bound-compression}}

\end{figure}

We call this method of producing approximate eigenvectors for $H_{QC}$
the compression method. This is because $H_{\rho}$ equals $\Pi_{\rho}H_{qc}\Pi_{\rho}$
on the appropriate subspace, where $\Pi_{\rho}$ is the projector
onto the span of the canonical basis elements located within the circle
of radius $\rho$. By looking for the eigenvalue of $H_{\rho}$ closest
to every $\lambda$ in a grid, of increment $0.002$ between
$E=0.6$ and $E=6.2$, and increment $0.05$ outside that interval
down to $0$ and up to $7$, we find upper bounds on the distance
to the spectrum as shown in Figure~\ref{fig:Upper-bound-compression}. 

In this computation, it was necessary to use a radius as large as
$\rho=691$. This corresponds to working on a Hilbert space of dimension
$3,621,440$. As we need to work at many energy levels, this is a
slow computation, and it turns out it is not so accurate. Figure~\ref{fig:Upper-bound-compression}
seems to indicate that there is a moderate gap in the bulk spectrum
around $E=0.8$. We will see this is a mirage.

How we find approximate eigenvectors that are zero at the boundary
is of no consequence as long as we compute the eigenerror. We hope
that researchers in numerical linear algebra will look at this problem.
In this work, eigensolvers are regarded as a black box, and the results
are then manipulated to force zero at the boundary. It is expected
that adapting an eigensolver to only consider vectors that are zero
at and near the boundary will produce better results. This is out
of the scope of this study, which is to illuminate methods physicists
can immediately adopt.

\section{Bulk Spectrum via the Spectral Localizer}

The spectral localizer \cite{loringSB_even_dim_localizer,LoringPseudospectra}
is designed to produce joint approximate eigenvalues for multiple
observables. We will apply the spectral localizer to the observables
$H_{\rho}-\lambda I$, $\kappa X_{\rho}$ and $\kappa Y_{\rho}$,
where $X_{\rho}$ and $Y_{\rho}$ are the observables for position
in the truncated round Hilbert space of radius $\rho$, where $\kappa$
is a tuning parameter, and $\lambda=E$ is one of the energy levels with
which we are concerned. The spectral localizer in
this case is 
\[
L=\left[\begin{array}{cc}
H_{\rho}-\lambda I & \kappa X_{\rho}-i\kappa Y_{\rho}\\
\kappa X_{\rho}+i\kappa Y_{\rho} & -H_{\rho}+\lambda I
\end{array}\right]
\]
 and we expect to find an eigenvalue for $L$ near $0$. Thus we have
$\psi_{1}$ and $\psi_{2}$ with $\left\Vert \psi_{1}\right\Vert ^{2}+\left\Vert \psi_{2}\right\Vert ^{2}=1$
and 
\[
\left[\begin{array}{cc}
H_{\rho}-\lambda I & \kappa X_{\rho}-i\kappa Y_{\rho}\\
\kappa X_{\rho}+i\kappa Y_{\rho} & -H_{\rho}+\lambda I
\end{array}\right]\left[\begin{array}{c}
\psi_{1}\\
\psi_{2}
\end{array}\right]\approx\left[\begin{array}{c}
0\\
0
\end{array}\right].
\]
We then expect 
\[
H_{\rho}\psi_{j}\approx\lambda\psi_{j},
\]
\[
\kappa X_{\rho}\psi_{j}\approx0
\]
 and 
\[
\kappa Y_{\rho}\psi_{j}\approx0.
\]

There is mathematics \cite{LoringPseudospectra} predicting the accuracy
of these approximations.  However, these are generic estimates that don't take into account the quasiperiodicity
of $H_{QC}$, so they don't deliver very useful bounds. In any
case, we only care about the following. We rescale each $\psi_{j}$,
so set 
\[
\psi_{j}'=\frac{1}{\|\psi_{j}\|}\psi_{j},
\]
and work with the $\psi_{j}'$ with the smaller eigenerror 
\[
\left\Vert H_{\rho}\psi_{j}'-\lambda\psi_{j}'\right\Vert .
\]
Finally, we apply the a taper function and rescale again, 
\[
\psi=\frac{1}{\left\Vert \tau(R)\psi_{j}\right\Vert }\tau(R)\psi_{j}
\]
where $\tau$ is as in Equation~\ref{eq:taper_function} and $R=\sqrt{X_{\rho}^{2}+Y_{\rho}^{2}}$.
This then gives an upper bound on 
\begin{equation}
\left\Vert H_{\rho}\psi_{j}-\lambda\psi_{j}\right\Vert \label{eq:spectral_errs_localizer}
\end{equation}
on the distance from $\lambda$ to the spectrum of $H_{QC}$.

\begin{figure}
\includegraphics[bb=180bp 270bp 430bp 525bp,clip,width=2.6in]{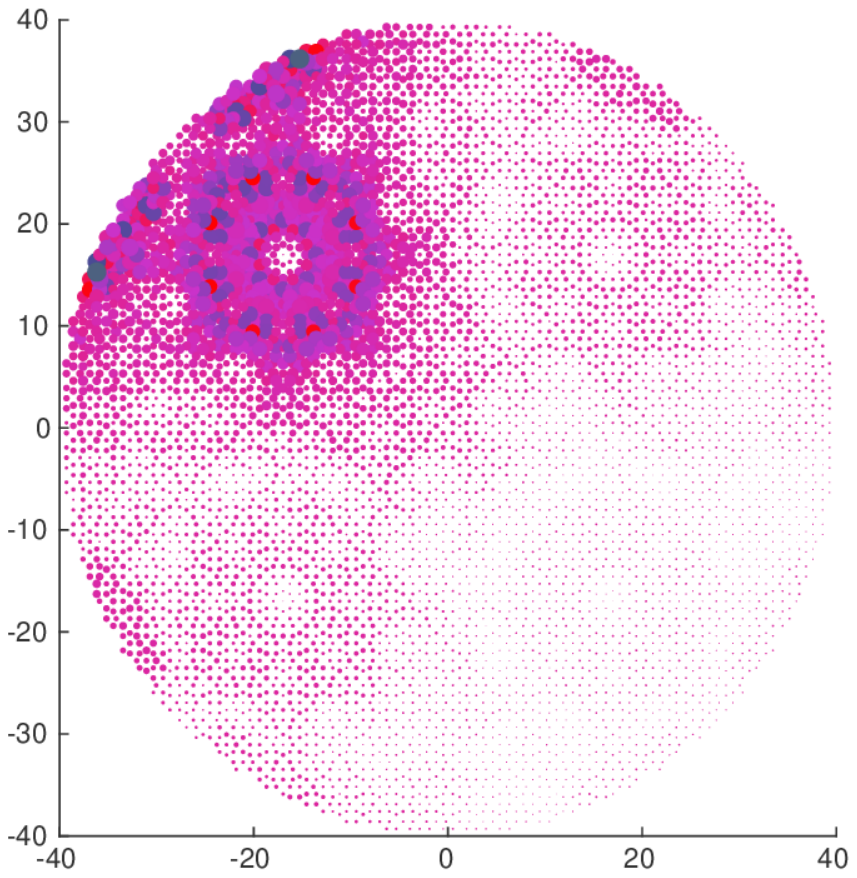}
\includegraphics[bb=180bp 270bp 430bp 525bp,clip,width=2.6in]{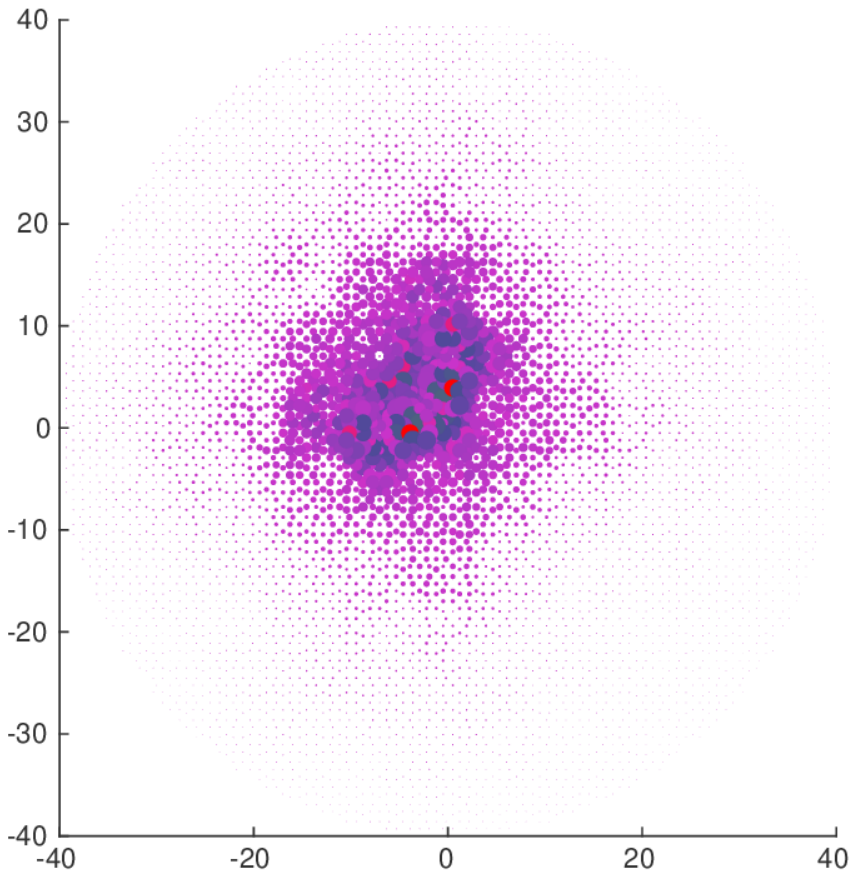}\\
\includegraphics[bb=180bp 270bp 430bp 525bp,clip,width=2.6in]{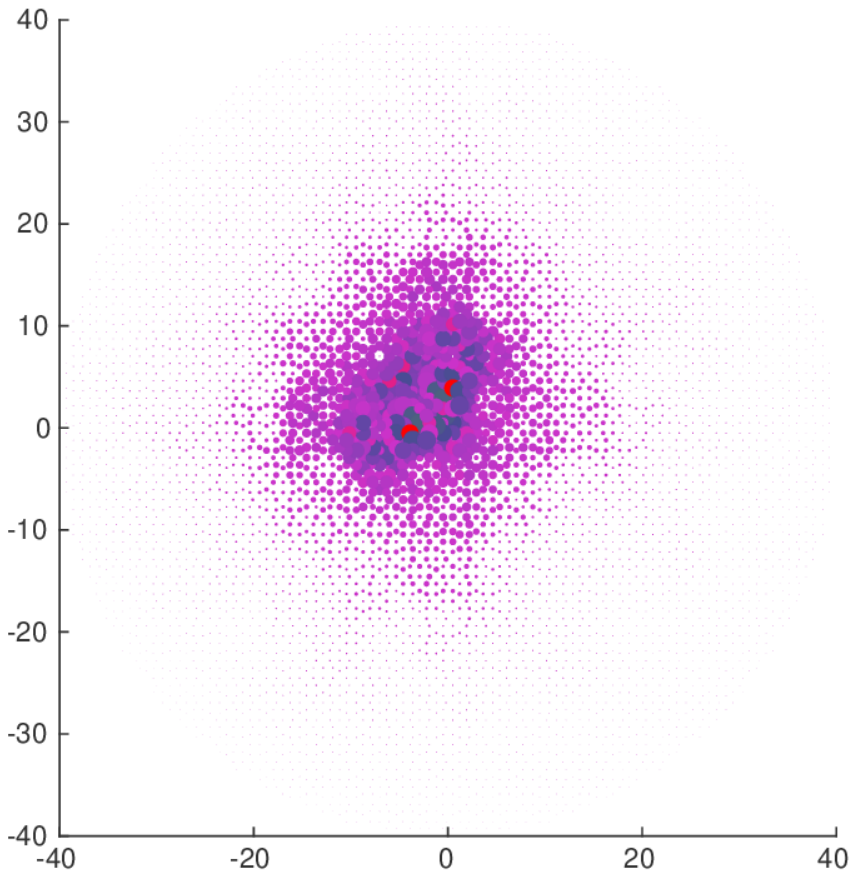}
\includegraphics[bb=180bp 270bp 430bp 525bp,clip,width=2.6in]{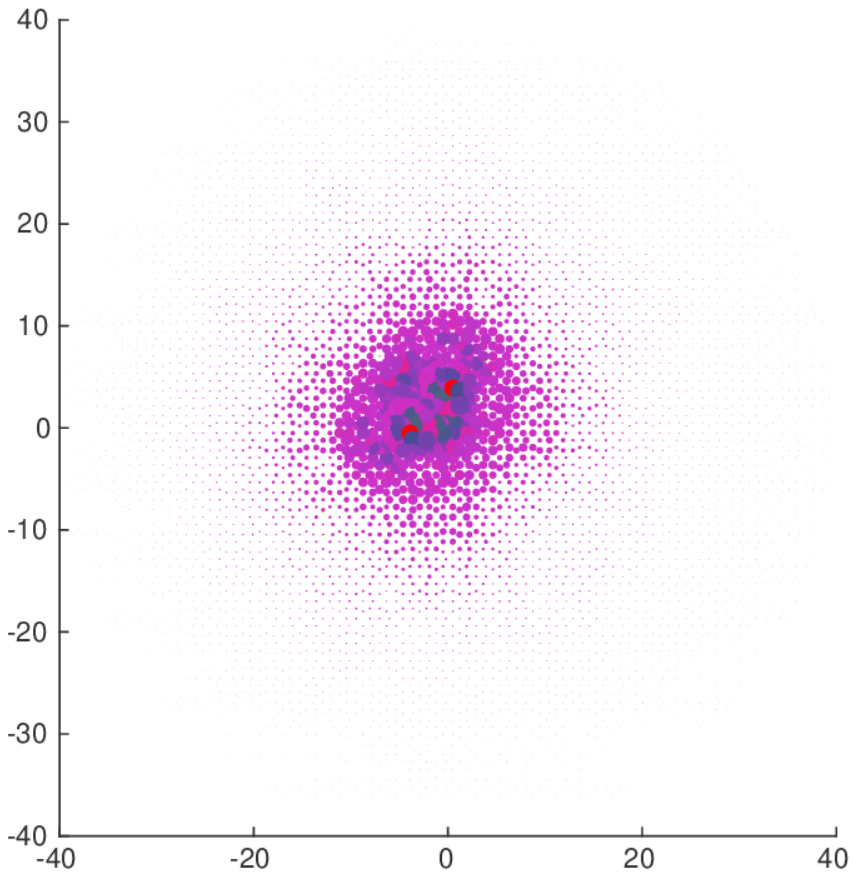}

\caption{States produced by the localizer method with increasing values of
$\kappa$ and with $\lambda$ set to $0.828.$ The top-left panel
has $\kappa=0.0002$. The top-right panel has $\kappa=0.0010$. The
bottom-left panel has $\kappa=0.0020$. The bottom-right panel has
$\kappa=0.0050$. \label{fig:Localizer-increasing_kappa}}

\end{figure}

\begin{figure}
\includegraphics{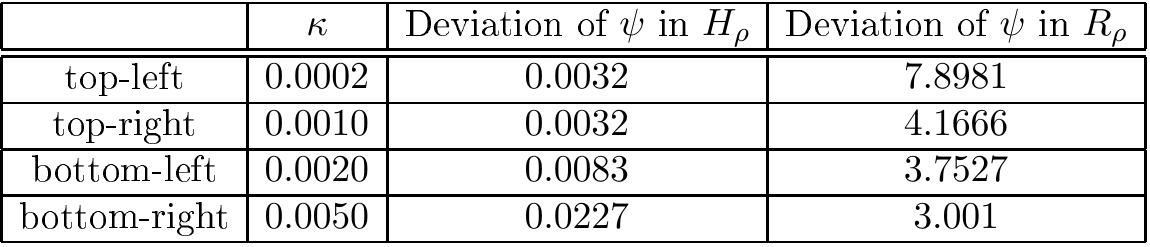}

\caption{Deviation of the states shown in Figure~\ref{fig:Localizer-increasing_kappa}
with respect to $H_{\rho}$ and to radius $R_{\rho}=\sqrt{X_{\rho}^{2}+Y_{\rho}^{2}}$.
\label{fig:Deviation-of-states}}

\end{figure}

If we set $\kappa=0$ then this method reduces to the compression method,
as both $\psi_{1}$ and $\psi_{2}$ will be exact eigenvectors for
$H_{QC}$ for an eigenvalue near $\lambda$. In the case of $\lambda=0.828$
and $\kappa=0$ and radius $40$, the results will be exactly the
state illustrated in Figure~\ref{fig:ambiguous_mode}. Figures~\ref{fig:Localizer-increasing_kappa} and \ref{fig:Deviation-of-states}
illustrate the effect on the state found for the same $\lambda$
and $\rho$ but with increasing values for $\kappa$.

\begin{figure}
\includegraphics[bb=180bp 270bp 430bp 525bp,clip,width=4in]{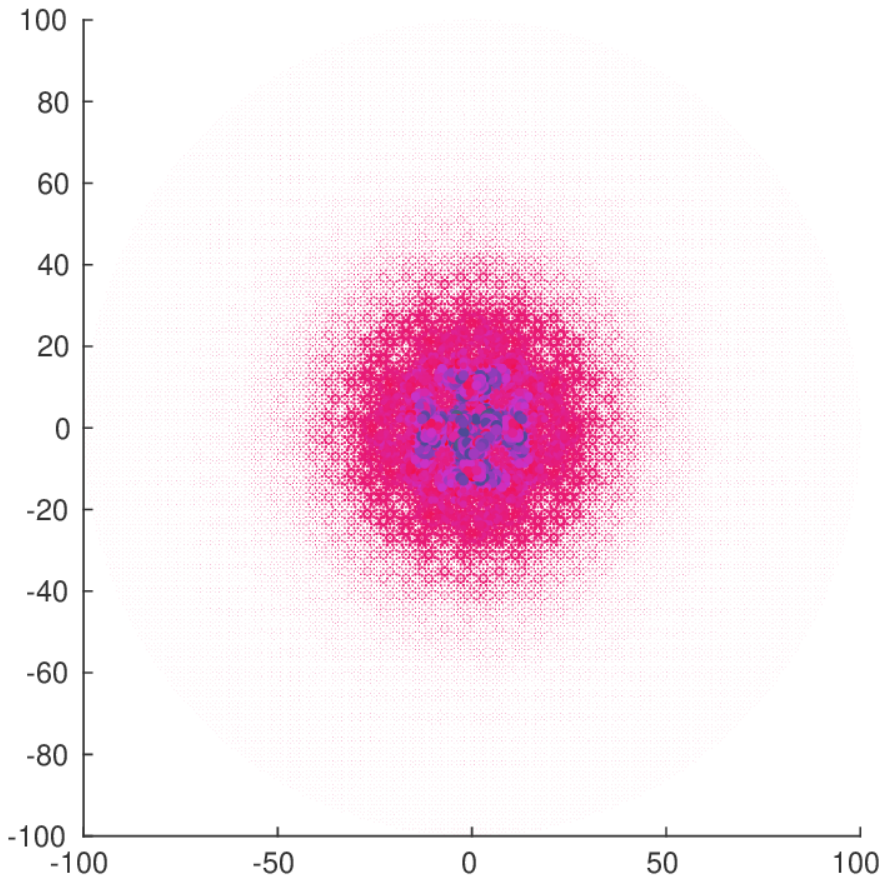}\\
\includegraphics[bb=180bp 270bp 430bp 525bp,clip,width=4in]{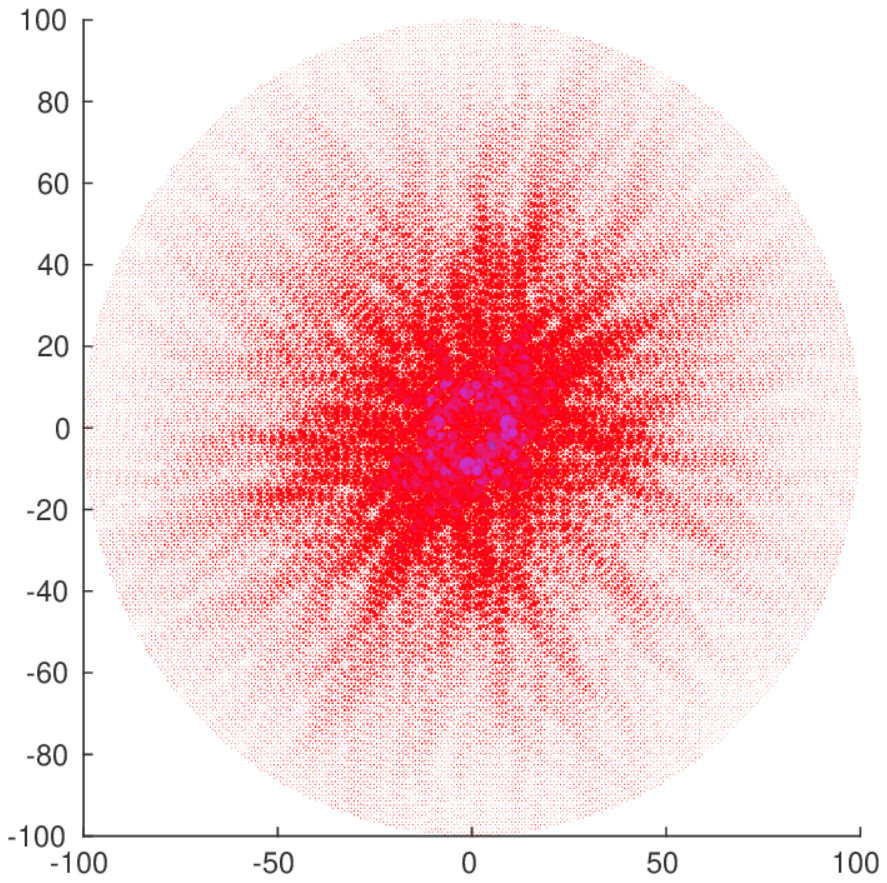}

\caption{Approximate eigenvalues for the finite system, radius $100$. Found
with the localizer method, with $\kappa=0.001$. On the top with approximate
eigenvalue $0.66$. On the bottom with approximate eigenvalue $4.8$.
\label{fig:bulk_states_approx}}

\end{figure}

With the correct setting of the tuning parameter, we expect to get
approximate eigenvectors for the finite Hamiltonian that are encouraged
to live away from the boundary. We hope that the loss of accuracy incurred
during the tapering process is enough smaller to compensate for the
fact that we are starting with only an approximate eigenvector. 

A few more of the approximate eigenvectors, before tapering, are shown
in Figure~\ref{fig:bulk_states_approx}. There are many energy levels
to explore, which the reader is encouraged to explore with the software
available as supplementary files \cite{Loringsupplement}.

\begin{figure}
\includegraphics[width=6in]{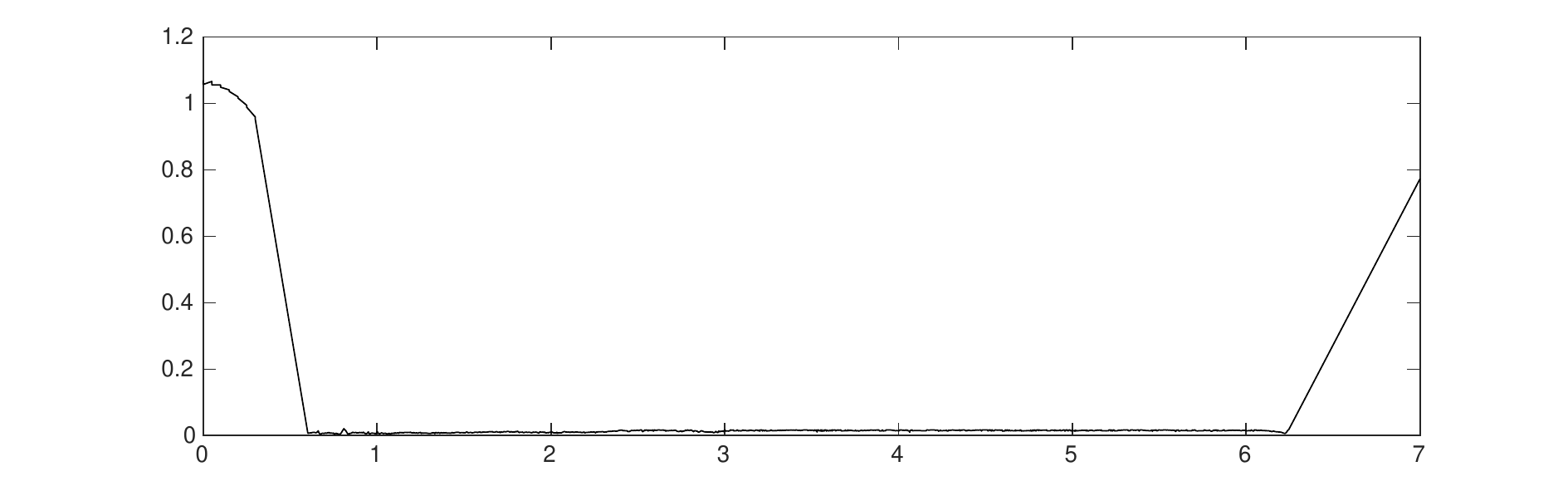}\\
\includegraphics[bb=0bp 0bp 544bp 170bp,width=6in]{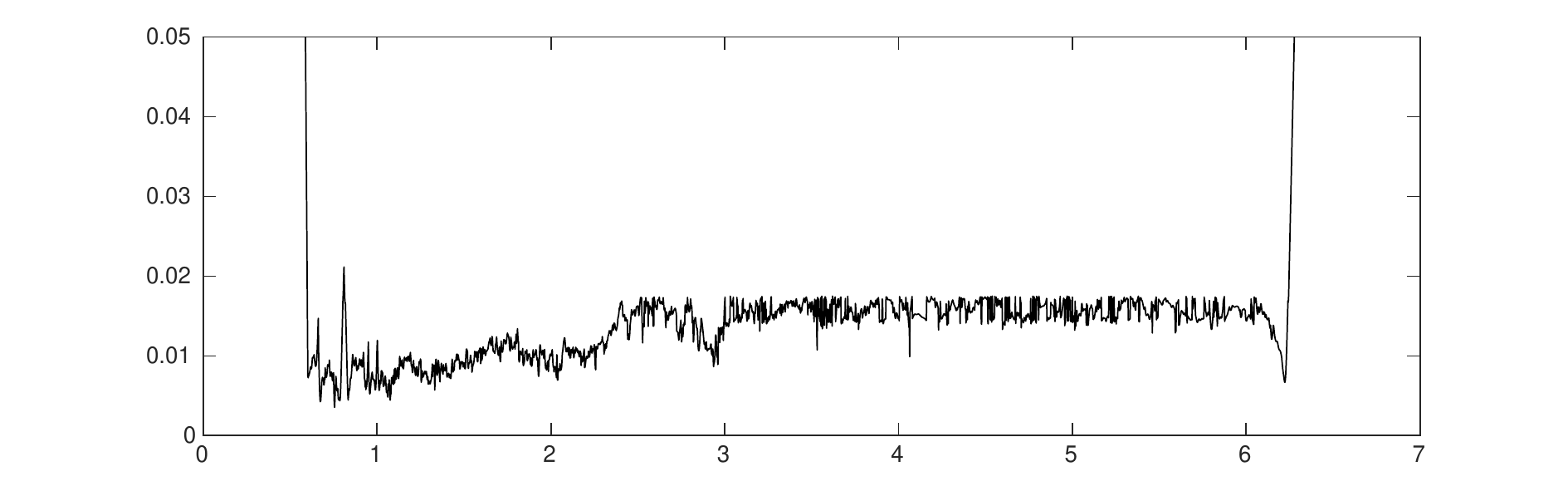}

\caption{The top figure shows the upper bound on the distance to the bulk spectrum via approximate eigenvalues
that were found by localizer method. The lower figure is zoomed in to clarify possible
small gaps. The local peak at $\lambda=0.081$ was computed as approximately
$0.021$. The left-most local minimum occurs at $\lambda=0.604$.
The right-most local minimum occurs at $\lambda=6.222$. \label{fig:Upper-bound-localizer} }
\end{figure}

We calculate upper bounds on the distance to the spectrum, on the
same grid of energy values as with the compression method. We refer
to this new method as the localizer method and the resulting upper
bound on the distance to the bulk spectrum is shown in Figure~\ref{fig:Upper-bound-localizer}.

We have thus computed the bulk spectrum to be contained in 
\[
[-6.227,-0.604]\cup[0.604,6.227]
\]
 and, if there is a gap within either of these bands, this gap will
be of diameter at most $0.044$. There is some evidence, as in Figure~\ref{fig:BigAndAmbiguous}
and Figure~\ref{fig:Upper-bound-localizer}, that there is a pair
of small gaps centered at about $\pm0.82$.

The claim that there is no large gap in $[0.604,6.227]$ is solid,
as its veracity rests only on the easy computation of the spectral
errors in Equation~\ref{eq:spectral_errs_localizer}. The claim that
the top of the spectrum is roughly $6.227$ is a little less rigorously
supported, as eigensolvers begin with a random seed and really bad
luck can cause then to report a smaller eigenvalue than what is possible.
The most speculative part of the claim is that the most inner eigenvalues
are close to $\pm0.604$. Making this a more rigorous claim will require
new results about the spectral localizer when used in quasiperiodic
conditions. It would make sense that someone do analytic work to get
a solid estimate of the radius of this gap, which is $\left\Vert H_{QC}^{-1}\right\Vert ^{-1}$.
The advantage of the present method is it runs in only a few days
on a computing cluster and should give an estimate in a reasonable
time for other non-crystalline systems.

\section{$K$-theory, Topological or $C^{*}$-Algebraic \label{sec:K-theory-Topological-or}}

There are two basic choices one can make when creating $K$-theoretical
indices for phases of matter \textendash{} to use topological $K$-theory
or to use the $K$-theory of $C^{*}$-algebras. The later is essentially
an extension of the former so it is more likely
to apply. The ring structure on $K^{*}(X)$ is lost when looking
at the $K$-theory of the $C^{*}$-algebra $C(X)$ but this is, so
far, not of consequence in topological physics. Most particularly,
when adding disorder it becomes more difficult to work with only topological
$K$-theory. This observation goes back to the work of Bellisard \cite{BellissardCstarSolidState}.
We have lost translation invariance for a different reason and so
also will utilize the $K$-theory of $C^{*}$-algebras.

As we have no symmetry of importance, we can work with the usual $C^{*}$-algebras,
which are algebras over $\mathbb{C}$ with a particular sort of norm.
The main examples are: the algebra $C(X)$ of continuous complex-valued
functions on a compact Hausdorff space $X$; the finite-dimensional
algebra $\boldsymbol{M}_{n}(\mathbb{C})$ for size $n$ square complex
matrices; and the Caulkin algebra $\mathbb{B}(\mathcal{H})/\mathbb{K}(\mathcal{H})$
of bounded operators on Hilbert space modulo the compact operators.
While $C(X)$ for $X$ a deformed sphere has a role in the mathematical
proofs \cite{loringSB_even_dim_localizer,loringSB_odd_dim_localizer}
that relate finite volume and infinite volume indices, here its only
role is motivational.

There are two groups associated to $A$, a $C^{*}$-algebra, denoted
$K_{0}(A)$ and $K_{1}(A)$. Technically there is also $K_{2}(A)$,
etc, but Bott periodicity \cite{RordamLandLbook} tells us $K_{2}(A)\cong K_{0}(A)$
and $K_{3}(A)\cong K_{1}(A)$, etc. It is also true that for a compact
Hausdorff space $X$ there is an isomorphism $K_{j}(C(X))\cong K^{j}(X)$,
so almost anything one can accomplish with topological $K$-theory
can just as well be accomplished using the $K$-theory of $C^{*}$-algebras.

We first look at $K_{1}(A)$ for a $C^{*}$-algebra $A$ that we assume
is unital. The usual picture of this in mathematics is built up from
homotopy classes of unitary elements in $A$ and the matrices over
$A$. An example, far older than the creation of $K$-theory, states
that there is an isomorphism
\[
\Psi:K_{1}\left(\mathbb{B}(\mathcal{H})/\mathbb{K}(\mathcal{H})\right)\rightarrow\mathbb{Z}
\]
which is based on the index of a Fredholm operator that is essentially
unitary:
\[
\Psi(\left[T+\mathbb{K}(\mathcal{H})\right])=\mathrm{index}(T).
\]

Because we are using complex scalars, our other example just ends
up with $K_{1}\left(\boldsymbol{M}_{n}(\mathbb{C})\right)=0$. This
reflects the elementary fact that any two unitary matrices of the
same size can be connected by a continuous path of unitary matrices. 

(Were we looking at real scalars, we would confront the fact that
two real orthogonal matrices can be connected by a continuous path
of real orthogonal matrices if and only if they have the same determinant.
See \cite{HastLorTheoryPractice,LoringPseudospectra} on how some
algorithms to compute indices of disordered 3D systems in class AII
reduce to calculating the sign of the determinant of a large matrix.)

The classical construction of $K_{0}(A)$, again with $A$ a unital
$C^{*}$-algebra, is in terms of projection in $A$ and the matrices
over $A$. A familiar example involves the $C^{*}$-algebra 
\[
C(\mathrm{T}^{2},\boldsymbol{M}_{n})
\]
of continuous functions from the torus to the $n$-by-$n$ matrices.
A projection $P$ is required, by definition, so be self-adjoint with
$P^{2}=P$. In this example, this means $P$ is a continuous function
\[
P:\mathbb{T}^{2}\rightarrow\boldsymbol{M}_{n}
\]
so that $P((\varphi,\theta))$ is the projector onto some continuously
varying subspace of $\mathbb{C}^{n}$. This is exactly what one finds
when looking at the Fermi projector in momentum space, when momentum
space exists.

The example closer to the work here is $K_{0}\left(\boldsymbol{M}_{n}(\mathbb{C})\right)$.
Any two projections in $\boldsymbol{M}_{n}(\mathbb{C})$ with the
same multiplicity for $1$ are homotopic. Since we are allowed matrices
over $\boldsymbol{M}_{n}(\mathbb{C})$, so $\boldsymbol{M}_{k}(\boldsymbol{M}_{n}(\mathbb{C}))$
which equals $\boldsymbol{M}_{kn}(\mathbb{C})$, we can achieve any
positive multiplicity. We get an isomorphism
\[
\Psi:K_{0}\left(\boldsymbol{M}_{n}(\mathbb{C})\right)\rightarrow\mathbb{Z}
\]
which, for projection $P$, is defined by
\[
\Psi\left([P]\right)=\dim(P(\mathbb{C}^{n})).
\]
Of course, this is just the multiplicity of the eigenvalue $1$. To
obtain a group we need ``negative multiplicity'' which arrives via
the Grotendieck group construction that arguments the data we have
with formal differences. So the full definition of $\Psi$ is 
\[
\Psi\left([P]-[Q]\right)=\dim(P(\mathbb{C}^{n}))-\dim(P(\mathbb{C}^{n})).
\]

This traditional picture of $K_{0}$ and $K_{1}$ is elegant, but
it is not ideal for numerical work. In practice, one comes across
a Hermitian matrix with a gap in its spectrum, and to apply the traditional
picture of $K$-theory one must compute a projection associated to
that gap. This is computationally intense, and if one is to preserve
sparsity one needs to accept $P^{2}\approx P$ anyway \cite{benzi2013decay}.
For some purposes, it is preferable to build $K_{1}(A)$ out of homotopy
classes of invertible operators. For $K_{0}(A)$ we prefer to use
homotopy classes of invertible Hermitian operators. Thus we are not
using a gap at $\tfrac{1}{2}$ as one does implicitly when considering
projections, but work with a gap at zero. This is also more compatible
with various symmetries needed for real $K$-theory \cite{BoersLorUnitaryKtheory}
or when working with class AII systems \cite{HastLorTheoryPractice}.

In this alternative picture of $K_{0}(A)$, for $A$ a unital $C^{*}$-algebra,
we use homotopy classes of invertible Hermitian elements in $A$ and
in matrices over $A$. To define the isomorphism
\[
\Psi:K_{0}\left(\boldsymbol{M}_{n}(\mathbb{C})\right)\rightarrow\mathbb{Z}
\]
it helps to use the signature. The signature $\mathrm{sig}(X)$ equals
the number, counting with multiplicity, of positive eigenvalues of
the given Hermitian matrix $X$ minus the number of negative eigenvalues.
Then as long as $n$ is even, we define $\Psi$ by 
\[
\Psi\left([H]\right)=\frac{1}{2}\mathrm{sig}(H)
\]
whenever $H$ is Hermitian and invertible.

\section{Numerical $K$-Theory}

The main theorem in prior work with Schulz-Baldes \cite{loringSB_even_dim_localizer}
tells us that the $K$-theory of a sufficiently large finite round
system will equal that of the infinite system. For the infinite system
there is no problem defining the $K$-theory class, or Chern number.
On the infinite system we have the Hamiltonian $H_{QC}$ and position
operators $X$ and $Y$. The Chern number is simply the index of a
Fredholm operator 
\[
\mathrm{ind}(H_{QC},X,Y)=\mathrm{ind}\left(\Pi_{v}\left(\frac{X+iY}{\left|X+iY\right|}\right)\Pi_{v}+(I-\Pi_{v})\right)
\]
on the full Hilbert space, and where $\Pi_{v}$ is the spectral projector
of $H_{QC}$ corresponding to the valence band \cite{bellissard1994noncommutative}.
It is usually hopeless to use numerical algorithms to compute the
index of a Fredholm operator, but in this case we rely on the spectral
localizer and the previously derived estimates \cite{loringSB_even_dim_localizer}.

The version of in idex of the finite system we need here is based
on the spectral localizer \cite{LoringPseudospectra}. One forms the
localizer at energy zero, 
\[
L_{\rho,\kappa}=\left[\begin{array}{cc}
H_{\rho} & \kappa X_{\rho}-i\kappa Y_{\rho}\\
\kappa X_{\rho}+i\kappa Y_{\rho} & -H_{\rho}
\end{array}\right],
\]
and the index of the system $(H_{\rho},X_{\rho},Y_{\rho})$ is computed,
with $\kappa$ at an appropriate value, as
\[
\mathrm{ind}(H_{\rho},X_{\rho},Y_{\rho})=\frac{1}{2}\mathrm{sig}\left(L_{\rho,\kappa}\right).
\]
This index is more robust, in terms of proven immutability under disorder
up to a certain strength, when the localizer has a larger gap at the
origin. It is equivalent to an element in the $K_{0}$-group of the
$C^{*}$-algebra of bounded operators on the finite volume Hilbert
space, or equivalently $K_{0}(\boldsymbol{M}_{n}(\mathbb{C}))$, as
explained in Section~\ref{sec:K-theory-Topological-or}.

We are guaranteed the equality
\[
\mathrm{ind}(H_{\rho},X_{\rho},Y_{\rho})=\mathrm{ind}(H_{QC},X,Y),
\]
according to the estimates obtained with Schulz-Baldes \cite[Equation 5]{loringSB_even_dim_localizer},
when we have 
\[
\rho>\frac{24\left\Vert H_{QC}\right\Vert \left\Vert \left[H_{QC},X+iY\right]\right\Vert }{\left\Vert H_{QC}^{-1}\right\Vert ^{-2}}.
\]
 We can get the estimate 
\[
\left\Vert \left[H_{QC},X+iY\right]\right\Vert \approx3.583
\]
by the same easy method used to estimate $\left\Vert H_{QC}\right\Vert $.
We put in our estimates and find we need a radius of 
\[
\rho_{0}=1480.
\]
We must (see Equation 6 in the paper with Schulz-Baldes \cite{loringSB_even_dim_localizer}) select $\kappa_{0}$ in the range 
\[
\frac{2g}{\rho_{0}}<\kappa_{0}\leq\frac{\left\Vert H_{QC}^{-1}\right\Vert ^{-3}}{12\left\Vert H\right\Vert \left\Vert \left[H,X+iY\right]\right\Vert },
\]
 meaning 
\[
0.000817\leq\kappa_{0}\leq0.000821,
\]
 so we select 
\[
\kappa_{0}=0.0008175.
\]

A radius of $1480$ leads to a Hilbert space of dimension of around
$16.6$ million, so the localizer will be a matrix with over 33 million
rows and column. It is sparse, but it is still annoyingly large. This
is too large to compute on a single node of a computing cluster. Rather
than resort to multi-node matrix factoring, we make a more modest
numerical calculation and use a numerical computation of a homotopy
to deduce that the signature of the huge localizer equals that of
the more modest localizer. Thus we get a computer assisted calculation
(not a fully rigorous proof) of what was surmised in prior work with
Fulga and Piklin \cite{fulga2016aperiodic}: 
\[
\mathrm{ind}(H_{QC},X,Y)=-1.
\]

\begin{figure}
\includegraphics[width=5in]{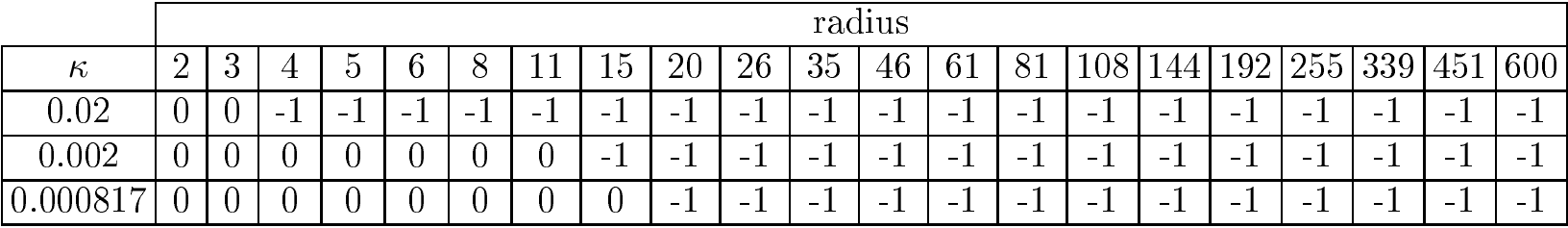}

\caption{Index for finite system of various radii and centers in the quasilattice.
These results are for a very small value of the tuning parameter,
$\kappa=0.0008175$, and a couple of larger values of $\kappa$. \label{fig:Index-varius-radii}}

\end{figure}

What can be computed today using Matlab on a computer with 64GB of
RAM is the signature of the localizer using radius of up to 600, which
corresponds to Hilbert space of dimension about $2.7$ million. While
not a proof, the data in Figure~\ref{fig:Index-varius-radii} is
more compelling than what has been computed using the older formulas,
such as the formula for the Bott index \cite{loring2011disordered}.
That index uses dense matrices, and cannot be used on today's computers
for systems with Hilbert space of dimension much larger than $20$
thousand. A common assumption about computing a Chern number is that
the ``various methods of computing it in real space'' are ``of similar
computational efficiency'' \cite{poyhonen2018amorphous} but this
seems to be false.

Notice that the data in Figure~\ref{fig:Index-varius-radii} is using
the small tuning parameter $\kappa_{0}$ and somewhat larger values.
For most purposes, a much larger value of $\kappa$ should be used,
as has been done in numerical studies \cite{fulga2016aperiodic,LoringPseudospectra}.
The very small value of the tuning parameter is used here to accommodate
the demands of the theorem we have \cite{loringSB_even_dim_localizer}
that relates finite systems to the infinite. The best way to set the
tuning parameter is illustrated in Figures \ref{fig:Localizer-increasing_kappa}
and \ref{fig:Deviation-of-states}. When a $\kappa$ value causes
the localizer to produce, on modest sizes systems, bulk states with
deviation in energy noticeably smaller than the interesting gaps in
the bulk, then that value of $\kappa$ should be valid for that system
size. Keeping $\kappa$ fixed, or varying as $C/\rho$, when moving
to larger systems leads to either a local or global $K$-theory index.

\begin{figure}
\includegraphics[bb=175bp 320bp 435bp 465bp,clip]{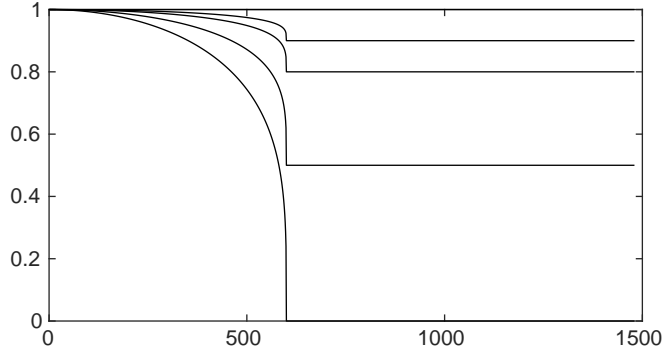}

\caption{The tapering function $f_{0}$ and points along the homotopy to the
constant function $f_{1}(r)=1$. Shown are $f_{t}$ for $t=0,0.5,0.8,0.9,1$.
\label{fig:The-tapering-function}}
\end{figure}

In order to relate a system of radius $\rho'=600$ to one of radius
$\rho_{0}=1480$, we consider two Hermitian matrices, $f_{0}(R_{\rho'})H_{\rho'}f_{0}(R_{\rho'})$
and $f_{0}(R_{\rho_{0}})H_{\rho_{0}}f_{0}(R_{\rho_{0}})$ where 
\[
f_{0}(r)=\begin{cases}
\left(1-\left(\frac{r}{\rho'}\right)^{2}\right)^{\frac{1}{4}} & \text{if }r<\rho'\\
0 & \text{if }r\geq\rho'
\end{cases}
\]
 and 
\[
R_{\rho}=\sqrt{X_{\rho}^{2}+Y_{\rho}^{2}}.
\]
 This function is the bottom one shown in Figure~\ref{fig:The-tapering-function}.
Since $f_{0}(R_{\rho_{0}})$ takes value zero beyond radius $\rho'$,
the localizer for the larger system devolves, on the Hilbert space
outside radius $\rho'$, into a direct sum of many matrices of the
form 
\[
\left[\begin{array}{cc}
0 & x-iy\\
x+iy & 0
\end{array}\right].
\]
 These have no effect on the signature, so 
\[
\mathrm{ind}(f_{0}(R_{\rho_{0}})H_{\rho_{0}}f_{0}(R_{\rho_{0}}),X_{\rho_{0}},Y_{\rho_{0}})=\mathrm{ind}(f_{0}(R_{\rho'})H_{\rho'}f_{0}(R_{\rho'}),X_{\rho'},Y_{\rho'}).
\]
 We can compute the latter index numerically and we find 
\[
\mathrm{ind}(f_{0}(R_{\rho'})H_{\rho'}f_{0}(R_{\rho'}),X_{\rho'},Y_{\rho'})=-1.
\]

To finish this argument we need to prove 
\begin{equation}
\mathrm{ind}(f_{0}(R_{\rho_{0}})H_{\rho_{0}}f_{0}(R_{\rho_{0}}),X_{\rho_{0}},Y_{\rho_{0}})=\mathrm{ind}(H_{\rho_{0}},X_{\rho_{0}},Y_{\rho_{0}})\label{eq:tapered_equals_untapered}
\end{equation}
 which we do using the homotopy invariance of $K$-theory. We define
\[
f_{t}(r)=(1-t)f_{0}(r)+t
\]
 as illustrated in Figure~\ref{fig:The-tapering-function}. Using
this path of functions we now continuously connect $f_{0}(R_{\rho_{0}})H_{\rho_{0}}f_{0}(R_{\rho_{0}})$
to $H_{\rho_{0}}$ via 
\[
f_{t}(R_{\rho_{0}})H_{\rho_{0}}f_{t}(R_{\rho_{0}}).
\]
If we can show that the resulting path of localizers 
\begin{equation}
\left[\begin{array}{cc}
f_{t}(R_{\rho_{0}})H_{\rho_{0}}f_{t}(R_{\rho_{0}}) & \kappa X_{\rho_{0}}-i\kappa Y_{\rho_{0}}\\
\kappa X_{\rho_{0}}+i\kappa Y_{\rho_{0}} & -f_{t}(R_{\rho_{0}})H_{\rho_{0}}f_{t}(R_{\rho_{0}})
\end{array}\right]\label{eq:path_of_localizers}
\end{equation}
remains invertible, we will be able to conclude that Equation~\ref{eq:path_of_localizers}
is true. This will then show 
\[
\mathrm{ind}(H_{\rho'},X_{\rho'},Y_{\rho'})=-1
\]
 and so 
\[
\mathrm{ind}(f_{0}(R_{\rho'})H_{\rho'}f_{0}(R_{\rho'}),X_{\rho'},Y_{\rho'})=-1.
\]
 That is, the infinite system is a Chern insulator.

\begin{figure}
\includegraphics[bb=175bp 325bp 430bp 470bp,clip]{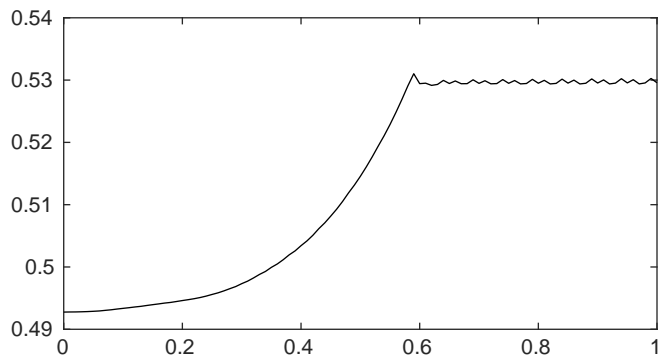}

\caption{Distance of the spectrum to $0$, computed at 101 evenly spaced points
along the path of localizers in Equation~\ref{eq:path_of_localizers}.
\label{fig:checking_homotopy}}

\end{figure}

Using analytic arguments, we are unable to show the path in Equation~\ref{eq:path_of_localizers}
is invertible. Such arguments \cite{loringSB_even_dim_localizer}
do work for paths of this form, but only for even larger systems.
Since the path is at every point Hermitian, we can use Weyl's estimate
on spectral variation and a calculation at a fine grid of points that
show the eigenvalue closest to $0$ is never very close to zero. We
display the results of such a computation in Figure~\ref{fig:checking_homotopy}.

\begin{figure}
\includegraphics[bb=180bp 355bp 430bp 440bp,clip]{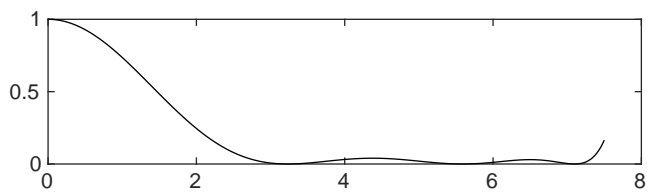}

\caption{The polynomial applied to the localizer in the power method to estimate
the innermost eigenvalue. \label{fig:The-polynomial-applied}}

\end{figure}

Each of the 101 points making the curve in Figure~\ref{fig:checking_homotopy}
was the result of estimating the eivenvalue closest to $0$ for an
$N$-by-$N$ matrix $L$ where $N$ is approximately $33$ million.
This is beyond the reach of basic eigensolvers. What was used here
was the basic power method applied to a polynomial in $L$. The polynomial
used is shown in Figure~\ref{fig:The-polynomial-applied}. This polynomial
was selected to do a good job estimating, for Hermitian matrices of
norm at most $7.4359$, the innermost eigenvalue if its absolute value
is in the range $[0.1,1.5]$. 

The implementation of this power method avoids matrix-matrix multiplication,
working only with matrix-vector multiplication. This is rather standard
numerical linear algebra, similar to some of the methods \cite{lin2016approximating}
used to estimate density of states when working with very large,
sparse matrices. One expects a researcher trained in numerical linear
algebra could do better, perhaps in the near future allowing us to
determine the $K$-theory of an infinite quasicrystalline system in
3D. 

\section{Discussion}

There are at least two notions of a local Chern number. There is the
method \cite{LoringPseudospectra} based on the spectral localizer
or Clifford pseudospectrum. To make this truly a local marker we need
to keep $\kappa$ rather large or fixed while the system size grows.
Probably the first local Chern marker was introduce by Kitaev \cite{kitaev2006anyons}
and was generalized by Bianco and Resta \cite{bianco2011mapping}. 

A local Chern number can be used to differentiate bulk spectrum from
edge spectrum in a finite sample. This strategy was employed in the
study of a Chern insulator in an amorphous system \cite{mitchell2018amorphous}.
Three-dimensional amorphous system appear to be capable of supporting
topological behavior \cite{agarwala2017topological} many dimensions
and symmetry classes which work with open systems, it should be possible
to verify that this 3D amorphous system has nontrivial $\mathbb{Z}_{2}$
index, perhaps using available software \cite{LoringPseudospectraMatlabCode}.

Much work \cite{ge2017topological,toniolo2017equivalence} has gone
into comparing the various ways to compute a (global) Chern number.
It seems that now would be a good time for someone to systematically
compare the values given by, and the computational complexity of, the various
means of defining a local $K$-theory marker. 

The reader might have expected a study involving larger finite samples
of the Floquet topological insulator introduced by Bandres, Rechtsman
and Segev \cite{bandres2016topological}. What makes that system more
challenging computationally is that a sparse time-dependent Hamiltonian
could well lead to a dense Floquet Hamiltonian. Without sparsity,
the localizer method for computing $K$-theory is not much more efficient
than using the Bott index. It may be that good sparse approximations
to the Floquet Hamiltonian exist, but calculating these is another
challenge in the realm of numerical linear algebra.

It would be nice to approximately compute the density of states for
$H_{QC}$ and related Hamilontians defined in terms of an infinite
quasicrystal. It seems that the spectral localizer method and the
methods used \cite{lin2016approximating} to approximate the density
of states for large but finite systems could be combined for this
purpose. Alternately, one could try to prove that the local density
of states for finite systems converges to something meaningful for
an infinite system. Both paths are worth exploring, as a result on
the density of states could somewhat make up for the lack of the momentum
space band picture of topological insulators on quasilattices. 

\section{Supplementary Files}

There are both data files and programs, all in Matlab formats, available
as supplements \cite{Loringsupplement} to this paper. The data files
are large portions of the Ammann-Beenker quasilattice. The main program
implements the compression and localizer method and displays the associated
states.

\section*{Acknowledgments }

This material is based upon work supported by the National Science Foundation under Grant No.\ DMS 1700102 and  by the Center for Advance Research Computing at the University of New Mexico.

\bibliographystyle{amsplain}
\bibliography{qc_band}

\end{document}